\documentclass[12pt]{article}
\usepackage{amsmath}
\usepackage{amssymb}
\usepackage{graphicx}
\usepackage{axodraw}
\usepackage{epsfig}
\usepackage{url}
\usepackage[small]{caption}
\begin{document}
\baselineskip 0.6cm

\def\simgt{\mathrel{\lower2.5pt\vbox{\lineskip=0pt\baselineskip=0pt
           \hbox{$>$}\hbox{$\sim$}}}}
\def\simlt{\mathrel{\lower2.5pt\vbox{\lineskip=0pt\baselineskip=0pt
           \hbox{$<$}\hbox{$\sim$}}}}
\def\simprop{\mathrel{\lower3.0pt\vbox{\lineskip=1.0pt\baselineskip=0pt
             \hbox{$\propto$}\hbox{$\sim$}}}}
\def\apparatusneut{\begin{picture}(15,15)
  \CArc(7.5,3)(6.5,0,360)
  \put(4.5,0){--}
\end{picture}}
\def\apparatusup{\begin{picture}(15,15)
  \CArc(7.5,3)(6.5,0,360)
  \put(4.5,0){$\uparrow$}
\end{picture}}
\def\apparatusdown{\begin{picture}(15,15)
  \CArc(7.5,3)(6.5,0,360)
  \put(4.5,0){$\downarrow$}
\end{picture}}
\def\apparatusone{\begin{picture}(15,15)
  \CArc(7.5,3)(6.5,0,360)
  \put(4.5,-0.7){$1$}
\end{picture}}
\def\apparatustwo{\begin{picture}(15,15)
  \CArc(7.5,3)(6.5,0,360)
  \put(4.5,-0.7){$2$}
\end{picture}}
\def\apparatusalpha{\begin{picture}(15,15)
  \CArc(7.5,3)(6.5,0,360)
  \put(4.1,0.2){$\alpha$}
\end{picture}}
\def\apparatusi{\begin{picture}(15,15)
  \CArc(7.5,3)(6.5,0,360)
  \put(5.2,-0.8){$i$}
\end{picture}}
\def\apparatusx{\begin{picture}(15,15)
  \CArc(7.5,3)(6.5,0,360)
  \put(4.4,0.2){$x$}
\end{picture}}
\def\apparatusy{\begin{picture}(15,15)
  \CArc(7.5,3)(6.5,0,360)
  \put(4.4,0.2){$y$}
\end{picture}}
\def\desk{\begin{picture}(15,15)
  \Line(3,8)(12,8) \Line(3,8)(0,3) \Line(12,8)(15,3) \Line(0,3)(15,3)
  \Line(1,3)(1,-3) \Line(14,3)(14,-3)
  \Line(3,3)(3,-1) \Line(12,3)(12,-1)
\end{picture}}
\def\obsbefore{\begin{picture}(15,15)
  \CArc(7.5,7.5)(2.5,0,360)
  \put(6.7,4.9){$\cdot$} \Line(10.0,7.8)(10.4,6.6) \Line(8.4,6.2)(9.5,6.0)
  \Line(7.5,4.5)(3,1.5) \Line(7.5,4.5)(12,1.5) \Line(7.5,4.5)(7.5,0.5)
  \Line(7.5,0.5)(5,-5) \Line(7.5,0.5)(10,-5)
\end{picture}}
\def\obsafter{\begin{picture}(15,15)
  \CArc(7.5,7.5)(2.5,0,360)
  \put(5.2,4.9){$\cdot$} \Line(5.0,7.8)(4.6,6.6) \Line(6.6,6.2)(5.5,6.0)
  \Line(7.5,4.5)(3,1.5) \Line(7.5,4.5)(12,1.5) \Line(7.5,4.5)(7.5,0.5)
  \Line(7.5,0.5)(5,-5) \Line(7.5,0.5)(10,-5)
\end{picture}}
\def\brainup{\begin{picture}(20,15)
  \Line(-3,9)(16,9) \CArc(16,7)(2,0,90) \Line(18,7)(18,-1.2)
  \CArc(-3,5)(4,0,90) \Line(1,5)(1,-1.2)
  \CArc(3,-1.2)(2,180,270) \Line(3,-3.2)(16,-3.2) \CArc(16,-1.2)(2,270,360)
  \CArc(9.5,5.5)(2.8,0,360) \put(7.6,5.6){\tiny ${}_\uparrow$}
  \Line(5.9,5.2)(6.7,5.2) \Line(12.3,5.2)(13.1,5.2)
  \Line(5.9,5.4)(3.5,1.4) \Line(13.1,5.4)(15.5,1.4) \Line(3.5,1.5)(15.5,1.5)
  \Line(4.3,1.5)(4.3,-2.7) \Line(14.7,1.5)(14.7,-2.7)
  \Line(5.9,1.5)(5.9,-1.3) \Line(13.1,1.5)(13.1,-1.3)
\end{picture}}
\def\braindown{\begin{picture}(20,15)
  \Line(-3,9)(16,9) \CArc(16,7)(2,0,90) \Line(18,7)(18,-1.2)
  \CArc(-3,5)(4,0,90) \Line(1,5)(1,-1.2)
  \CArc(3,-1.2)(2,180,270) \Line(3,-3.2)(16,-3.2) \CArc(16,-1.2)(2,270,360)
  \CArc(9.5,5.5)(2.8,0,360) \put(7.6,5.6){\tiny ${}_\downarrow$}
  \Line(5.9,5.2)(6.7,5.2) \Line(12.3,5.2)(13.1,5.2)
  \Line(5.9,5.4)(3.5,1.4) \Line(13.1,5.4)(15.5,1.4) \Line(3.5,1.5)(15.5,1.5)
  \Line(4.3,1.5)(4.3,-2.7) \Line(14.7,1.5)(14.7,-2.7)
  \Line(5.9,1.5)(5.9,-1.3) \Line(13.1,1.5)(13.1,-1.3)
\end{picture}}
\def\brainalpha{\begin{picture}(20,15)
  \Line(-3,9)(16,9) \CArc(16,7)(2,0,90) \Line(18,7)(18,-1.2)
  \CArc(-3,5)(4,0,90) \Line(1,5)(1,-1.2)
  \CArc(3,-1.2)(2,180,270) \Line(3,-3.2)(16,-3.2) \CArc(16,-1.2)(2,270,360)
  \CArc(9.5,5.5)(2.8,0,360) \put(7.3,5.6){\tiny ${}_\alpha$}
  \Line(5.9,5.2)(6.7,5.2) \Line(12.3,5.2)(13.1,5.2)
  \Line(5.9,5.4)(3.5,1.4) \Line(13.1,5.4)(15.5,1.4) \Line(3.5,1.5)(15.5,1.5)
  \Line(4.3,1.5)(4.3,-2.7) \Line(14.7,1.5)(14.7,-2.7)
  \Line(5.9,1.5)(5.9,-1.3) \Line(13.1,1.5)(13.1,-1.3)
\end{picture}}
\def\brainright{\begin{picture}(20,15)
  \Line(-3,9)(16,9) \CArc(16,7)(2,0,90) \Line(18,7)(18,-1.2)
  \CArc(-3,5)(4,0,90) \Line(1,5)(1,-1.2)
  \CArc(3,-1.2)(2,180,270) \Line(3,-3.2)(16,-3.2) \CArc(16,-1.2)(2,270,360)
  \Line(6.5,2.7)(12.5,2.7) \Line(6.5,8.3)(12.5,8.3)
  \Line(6.5,2.7)(6.5,8.3) \Line(12.5,2.7)(12.5,8.3)
  \put(5.8,5.3){\tiny ${}_\rightarrow$}
  \Line(5.9,5.2)(6.5,5.2) \Line(12.5,5.2)(13.1,5.2)
  \Line(5.9,5.4)(3.5,1.4) \Line(13.1,5.4)(15.5,1.4) \Line(3.5,1.5)(15.5,1.5)
  \Line(4.3,1.5)(4.3,-2.7) \Line(14.7,1.5)(14.7,-2.7)
  \Line(5.9,1.5)(5.9,-1.3) \Line(13.1,1.5)(13.1,-1.3)
\end{picture}}
\def\brainleft{\begin{picture}(20,15)
  \Line(-3,9)(16,9) \CArc(16,7)(2,0,90) \Line(18,7)(18,-1.2)
  \CArc(-3,5)(4,0,90) \Line(1,5)(1,-1.2)
  \CArc(3,-1.2)(2,180,270) \Line(3,-3.2)(16,-3.2) \CArc(16,-1.2)(2,270,360)
  \Line(6.5,2.7)(12.5,2.7) \Line(6.5,8.3)(12.5,8.3)
  \Line(6.5,2.7)(6.5,8.3) \Line(12.5,2.7)(12.5,8.3)
  \put(6.6,5.3){\tiny ${}_\leftarrow$}
  \Line(5.9,5.2)(6.5,5.2) \Line(12.5,5.2)(13.1,5.2)
  \Line(5.9,5.4)(3.5,1.4) \Line(13.1,5.4)(15.5,1.4) \Line(3.5,1.5)(15.5,1.5)
  \Line(4.3,1.5)(4.3,-2.7) \Line(14.7,1.5)(14.7,-2.7)
  \Line(5.9,1.5)(5.9,-1.3) \Line(13.1,1.5)(13.1,-1.3)
\end{picture}}
\def\apparatusxneut{\begin{picture}(15,15)
  \Line(1,-3)(14,-3) \Line(1,9)(14,9)
  \Line(1,-3)(1,9) \Line(14,-3)(14,9)
  \put(4.5,0){--}
\end{picture}}
\def\apparatusxright{\begin{picture}(15,15)
  \Line(1,-3)(14,-3) \Line(1,9)(14,9)
  \Line(1,-3)(1,9) \Line(14,-3)(14,9)
  \put(1.5,0){$\rightarrow$}
\end{picture}}
\def\apparatusxleft{\begin{picture}(15,15)
  \Line(1,-3)(14,-3) \Line(1,9)(14,9)
  \Line(1,-3)(1,9) \Line(14,-3)(14,9)
  \put(1.5,0){$\leftarrow$}
\end{picture}}
\def\ftapparatusneut{\begin{picture}(12,12)
  \CArc(6,2.4)(5.2,0,360)
  \put(3.6,0){--}
\end{picture}}
\def\bra#1{\left< #1 \right|}
\def\ket#1{\left| #1 \right>}
\def\inner#1#2{\left< #1 | #2 \right>}

\begin{titlepage}

\begin{flushright}
UCB-PTH-11/08 \\
\end{flushright}

\vskip 1.3cm

\begin{center}
{\Large \bf Quantum Mechanics, Spacetime Locality, and Gravity}

\vskip 0.7cm

{\large Yasunori Nomura}

\vskip 0.4cm

{\it Berkeley Center for Theoretical Physics, Department of Physics,\\
 University of California, Berkeley, CA 94720, USA}

\vskip 0.1cm

{\it Theoretical Physics Group, Lawrence Berkeley National Laboratory,
 CA 94720, USA}

\vskip 0.8cm

\abstract{Quantum mechanics introduces the concept of probability to 
 physics at the fundamental level, yielding {\it the measurement problem}. 
 On the other hand, recent progress in cosmology has led to the 
 ``multiverse'' picture, in which our observed universe is only one 
 of the many, bringing an apparent arbitrariness in defining 
 probabilities, called {\it the measure problem}.

 In this paper, we discuss how these two problems are intimately related 
 with each other, developing a complete picture for quantum measurement 
 and cosmological histories in the quantum mechanical universe.  On 
 one hand, quantum mechanics eliminates the arbitrariness of defining 
 probabilities in the multiverse, as discussed previously in arXiv:1104.2324. 
 On the other hand, the multiverse allows for understanding why we 
 observe an ordered world obeying consistent laws of physics, by providing 
 an {\it infinite-dimensional} Hilbert space.  This results in the 
 irreversibility of quantum measurement, despite the fact that the 
 evolution of the multiverse state is unitary.

 In order to describe the cosmological dynamics correctly, we need to 
 identify the structure of the Hilbert space for a system with gravity. 
 We argue that in order to keep spacetime locality in the description, 
 which plays a crucial role in understanding quantum measurement, the 
 Hilbert space for dynamical spacetime must be defined only in restricted 
 spacetime regions:\ in and on the (stretched) apparent horizon {\it as 
 viewed from a fixed reference frame}.  This requirement arises from 
 fixing/eliminating all the redundancies and overcountings in a general 
 relativistic, global spacetime description of nature.  It is responsible 
 for horizon complementarity as well as the ``observer dependence'' 
 of horizons/spacetime---these phenomena arise because changes of 
 the reference frame are represented in the Hilbert space defined on 
 restricted spacetime regions.  This can be viewed as an extension of 
 the Lorentz/Poincar\'{e} transformation in the quantum gravitational 
 context, as the Lorentz transformation is viewed as an extension of 
 the Galilean transformation.

 Given an initial condition, the evolution of the multiverse state 
 obeys the laws of quantum mechanics---it evolves deterministically 
 and unitarily, asymptoting to the ``heat death'' consisting of 
 supersymmetric Minkowski and singularity states.  The beginning 
 of the multiverse, however, is still an open issue.}

\newpage

\tableofcontents

\end{center}
\end{titlepage}

\section{Introduction---The Basic Picture}
\label{sec:intro}

This paper discusses two subjects:\ quantum mechanics and gravity, 
especially in the context of cosmology.  Quantum mechanics introduced 
the concept of probability to physics {\it at the fundamental level}. 
This has led to the issue of the quantum-to-classical transition, in 
particular the {\it measurement} problem.  Despite much progress, a 
complete and satisfactory picture, particularly the one including the 
entire universe, still seems missing.

Recent progress in cosmology has led to the ``multiverse'' picture---our 
observed universe may be one of the many in which low energy physical 
laws take different forms.  This view is suggested by both observation 
and theory:\ it provides a successful understanding of the order of 
magnitude of the observed dark energy~\cite{Weinberg:1987dv}, and 
arises naturally as a result of eternal inflation~\cite{Guth:1982pn} 
and the string landscape~\cite{Bousso:2000xa}.  This elegant picture, 
however, suffers from the issue of predictivity---in the multiverse, 
any event that can happen will happen in infinitely many times (due to 
eternally expanding spacetime), making any definition of probabilities 
extremely subtle~\cite{Guth:2000ka}.  Many proposals have been put 
forward to regulate these infinities, but they seem to be arbitrary, 
without relying on a solid fundamental principle.  This arbitrariness 
of defining probabilities in the multiverse is called the {\it measure} 
problem, and has been a focus of much recent research.

More recently, it has been proposed that the above two issues---both 
connected to probabilities---are in fact related~\cite{Nomura:2011dt}. 
In particular, the probabilities in the eternally inflating multiverse 
must be defined based on the principles of quantum mechanics, which 
eliminates the ambiguity for the definition (as well as the problems 
and paradoxes plaguing some of the earlier measures).  The probability 
formula given in Ref.~\cite{Nomura:2011dt} takes the form
\begin{equation}
  P(B|A) = \frac{\int\!dt \bra{\Psi(t)} {\cal O}_{A \cap B} \ket{\Psi(t)}} 
    {\int\!dt \bra{\Psi(t)} {\cal O}_A \ket{\Psi(t)}},
\label{eq:probability-AB}
\end{equation}
where $\ket{\Psi(t)}$ is the state representing the entire multiverse, 
while ${\cal O}_A$ and ${\cal O}_{A \cap B}$ are projection operators 
implementing physical questions one would ask.  This is (essentially) 
the Born rule.  Indeed, the formula of Eq.~(\ref{eq:probability-AB}) 
can be used to answer questions both regarding global properties of 
the universe and outcomes of particular experiments, providing complete 
unification of the eternally inflating multiverse and many worlds 
in quantum mechanics.

In Ref.~\cite{Nomura:2011dt}, it was argued that the state $\ket{\Psi(t)}$ 
must be defined only in the restricted spacetime regions---in 
and on the (stretched) apparent horizons---consistently with 
what we learned about quantum gravity in the past two decades:\ 
the holographic principle~\cite{'tHooft:1993gx} and black hole 
complementarity~\cite{Susskind:1993if}.  In the cosmological context, 
however, the locations of horizons are ``observer dependent.''  What 
does this really mean?  Moreover, Ref.~\cite{Nomura:2011dt} also discussed 
the meaning of spacetime singularities from the low energy viewpoint, 
and argued that it implies that the multiverse evolves asymptotically 
into a supersymmetric Minkowski world.  Do these results have any 
implications for the problem of quantum measurement?

In this paper, we study these issues, developing a complete picture 
for quantum measurement and cosmological histories in the quantum 
mechanical universe.  A crucial ingredient for our discussion is the 
structure of the Hilbert space corresponding to semi-classical spacetime, 
which is identified in Ref.~\cite{Nomura:2011dt} (and will be suitably 
refined here):
\begin{equation}
  {\cal H} = \bigoplus_{\cal M} {\cal H}_{\cal M},
\qquad
  {\cal H}_{\cal M} = {\cal H}_{{\cal M}, {\rm bulk}} 
    \otimes {\cal H}_{{\cal M}, {\rm horizon}},
\label{eq:multiverse-H}
\end{equation}
where ${\cal H}_{\cal M}$ is the Hilbert (sub)space for a set of fixed 
semi-classical geometries ${\cal M}$ that have the same (stretched) 
apparent horizon $\partial {\cal M}$, {\it as viewed from a local 
Lorentz frame of a fixed spatial point $p$}.  It consists of the parts 
corresponding to the regions in and on the horizon, ${\cal H}_{{\cal M}, 
{\rm bulk}}$ and ${\cal H}_{{\cal M}, {\rm horizon}}$, both of which 
have the dimension of $\exp({\cal A}_{\partial {\cal M}}/4)$, where 
${\cal A}_{\partial {\cal M}}$ is the area of the horizon in Planck units. 
(We will see that the complete Hilbert space for quantum gravity also 
has ``intrinsically quantum mechanical'' elements associated with 
spacetime singularities, but they are irrelevant for physical predictions.) 
We argue that, as quantum mechanics has helped the measure problem in 
eternal inflation, the multiverse helps the measurement problem in quantum 
mechanics, which demonstrates another ``cooperation'' between quantum 
mechanics and cosmology.  In particular, the fact that we observe 
an ordered, classical world is explained by a combination of 
spacetime locality {\it and} the fact that the multiverse ultimately 
evolves into a Minkowski (or singularity) world, which has an 
{\it infinite-dimensional} Hilbert space
\begin{equation}
  {\rm dim}\,{\cal H}_{\cal M} = \infty
\quad\mbox{for}\quad
  {\cal M} = {\rm Minkowski}.
\label{eq:Minkowski-inf}
\end{equation}
This results in the irreversibility of quantum measurement, despite the 
fact that the evolution of the multiverse state is unitary.

We also elucidate the meaning of the Hilbert space structure in 
Eq.~(\ref{eq:multiverse-H}), which provides a better understanding of 
quantum gravity.  It is well known that to do Hamiltonian quantum mechanics, 
all the gauge redundancies must be fixed---and a theory of gravity has 
huge redundancies associated with general coordinate transformations. 
Defining a state in Eq.~(\ref{eq:multiverse-H}) provides a simple way 
to fix these redundancies and to extract causal relations among events, 
which are physical (coordinate reparameterization invariant).  In other 
words, {\it we need to fix a reference frame when we describe a system 
with gravity quantum mechanically}---this is the real meaning of the 
phrase:\ ``physics must be described from the viewpoint of a single 
observer'' in Ref.~\cite{Nomura:2011dt}.  In particular, the location 
of a physical object/observer (with respect to ``the origin of 
the coordinates'' $p$) has physical meaning, so it needs to be 
included as a part of specification in condition $A$ when applying 
Eq.~(\ref{eq:probability-AB}).

Since the Hilbert space ${\cal H}$ in Eq.~(\ref{eq:multiverse-H}) 
is defined on restricted spacetime regions, changes of the reference 
frame represented in ${\cal H}$ in general mix elements of different 
${\cal H}_{\cal M}$ as well as the degrees of freedom associated with 
${\cal H}_{{\cal M}, {\rm bulk}}$ and ${\cal H}_{{\cal M}, {\rm horizon}}$. 
(More generally, changing the reference frame can also mix elements 
of Eq.~(\ref{eq:multiverse-H}) with intrinsically quantum mechanical 
states associated with singularities.)  This is the origin of horizon 
complementarity (mixture between different ${\cal H}_{\cal M}$) and 
of the ``observer dependence'' of cosmic horizons (mixture between 
the bulk and horizon degrees of freedom)!  This general transformation 
can be viewed as an extension of the Lorentz/Poincar\'{e} transformation 
in the quantum gravitational context.  It introduces more ``relativeness'' 
in physical descriptions---it makes even the concept of spacetime 
relative, as it mixes the bulk and horizon degrees of freedom in general.

Two key aspects of our picture of quantum measurement are dynamical 
evolution and the infinite dimensionality of the Hilbert space, given 
(partly) by Eqs.~(\ref{eq:multiverse-H},~\ref{eq:Minkowski-inf}).  We 
argue that spacetime locality---a special property of the time evolution 
operator---plays a crucial role in the evolution of a state.  It leads 
to ``amplification'' of classical information~\cite{q-Darwinism} {\it 
in a single component} of the multiverse state.  Schematically,
\begin{equation}
  \ket{\uparrow} \,\,\rightarrow\,\, 
    \ket{\uparrow} \bigl|\apparatusup\bigr> \,\rightarrow\, 
    \ket{\uparrow} \bigl|\apparatusup\bigr> \bigl|\obsafter\brainup\bigr> 
    \,\,\rightarrow\,\, \cdots,
\label{eq:amplification}
\end{equation}
which shows that the classical information (i.e.\ that the spin is up) 
is amplified in a detector pointer and brain state of an observer 
(which can be further amplified, e.g., in a note and a paper on the 
experiment, brain states of the people who have read the paper, and 
so on).  Note that since the faithful duplication of quantum information 
is prohibited~\cite{Wootters:1982zz}, only classical information can 
be amplified, whose content is much smaller than the full quantum 
information.%
\footnote{This does {\it not} mean that we cannot observe a superposition 
 of classically different configurations.  It just says that the 
 statement ``the system was in a superposition state'' is already 
 classical, and it is this information that is actually amplified.}
At the same time, the dynamical evolution also leads to 
``branching''~\cite{Everett:1957hd}:\ the state splits into many 
different components having well-defined classical configurations. 
For example, the initial $e^+ e^-$ state becomes a superposition 
of various components having well-defined particle configurations:
\begin{equation}
  \ket{e^+ e^-} \,\,\rightarrow\,\, 
  \ket{e^+ e^-} + \ket{\mu^+ \mu^-} + \cdots 
  + \ket{e^+ e^- e^+ e^-} + \cdots 
  \,\,\rightarrow\,\, \cdots,
\label{eq:branching}
\end{equation}
where we have omitted the coefficients for various components as well 
as momentum and spin indices.  Note that through this process, the same 
quantum information can be distributed into {\it multiple components} 
over time; what the no-cloning theorem prohibits is the duplication of 
quantum information {\it in a single component}.

The evolution of the multiverse state experiences both these effects 
as it evolves in the full quantum gravitational Hilbert space. 
Schematically,
\begin{equation}
  \ket{\Sigma} \,\,\rightarrow\,\, 
    \ket{A} + \ket{B} \,\,\rightarrow\,\, 
    \ket{aa} + \ket{bb} + \ket{cc} + \ket{dd} \,\,\rightarrow\,\, 
    \ket{\alpha\alpha\cdots\alpha} + \ket{\beta\beta\cdots\beta} + \cdots,
\label{eq:multiverse-evolve}
\end{equation}
where the various letters indicate classical information.  This evolution 
is deterministic and unitary, i.e., obeys the basic laws of quantum 
mechanics.  The amplification generically occurs from a smaller system 
to larger systems.  At the early stage of this process, the basis of the 
amplification is determined by the details of the system, as the standard 
analysis of decoherence shows~\cite{Schlosshauer}.  On the other hand, 
at later stages, where the relevant systems are large, the amplification 
occurs in the basis corresponding to states having well-defined classical 
configurations, as a result of spacetime locality. Various components 
of the state will then correspond to {\it different macroscopic worlds}, 
which will eventually evolve into different supersymmetric Minkowski 
(or singularity) states.  Since the Hilbert space dimension of Minkowski 
space is infinite, these worlds do not recohere---they really branch 
into different worlds!

The above picture provides a complete account for the process of quantum 
measurement in the eternally inflating multiverse.  While not all 
the aspects of the dynamics described above are fully proven, the basic 
picture is strongly supported by recent progress on understanding the 
quantum-to-classical transition (e.g.~\cite{q-Darwinism,Schlosshauer}). 
What are the implications of this in calculating physical probabilities 
in Eq.~(\ref{eq:probability-AB})?  Physical information we can handle 
is only the ``robust'' kind, i.e.\ the one that can appear multiple times 
in physical systems (e.g.\ as data stored in some ``memories,'' including 
someone's brain state).  It therefore only makes sense to ask questions 
about information that is amplified in some component of the state. 
This corresponds to choosing projection operators ${\cal O}_A$ and 
${\cal O}_{A \cap B}$ to extract only such information; in particular, 
it corresponds to projecting onto classically well-defined configurations 
when we ask questions about macroscopic systems.

The framework described here provides a solid theoretical ground for 
asking any physical questions in the quantum universe.  However, to 
make actual predictions in the context of the multiverse, e.g.\ of the 
value of a physical parameter we observe, we still need to know the 
explicit form of the time evolution operator as well as the initial 
condition for the multiverse state (except for a few special cases, 
including that for calculating the distribution of the cosmological 
constant~\cite{Larsen:2011mi}).  In particular, knowing the complete 
evolution of the state requires understanding of the dynamics of the 
horizon degrees of freedom as well as the full string landscape.  The 
former can be bypassed if we adopt the semi-classical approximation based 
on the ``bulk density matrix''~\cite{Nomura:2011dt}, $\rho_{\rm bulk}(t) 
= {\rm Tr}_{\rm horizon} \ket{\Psi(t)} \bra{\Psi(t)}$, while the latter 
needs further progress in string theory.  The initial condition for 
the multiverse state must be given by some external theory.  Some 
(speculative) possibilities are presented in Ref.~\cite{Nomura:2011dt}, 
but here we leave this issue aside and simply assume that an appropriate 
initial state is provided by some theory of initial conditions.

The organization of this paper is as follows.  In the first half of 
the paper, Sections~\ref{sec:interpret}~--~\ref{sec:QM-classical}, 
we discuss quantum measurement without taking into account the effect 
of gravity; readers who are only interested in quantum gravitational 
aspects might proceed directly to Section~\ref{sec:locality-grav}.  In 
Section~\ref{sec:interpret} we provide a pedagogical introduction to the 
problem of quantum measurement, especially the problem of the preferred 
basis.  In Section~\ref{sec:locality} we discuss carefully how spacetime 
locality selects a basis {\it in the Hilbert space}, a necessary 
ingredient to discuss basis selection {\it for quantum measurement}. 
In Section~\ref{sec:QM-classical} we analyze quantum measurement 
in the context of applying quantum mechanics to the whole universe. 
We argue that the preferred basis for measurement is determined purely 
by the dynamics; in particular, the ultimate openness of the system is 
not required.  Spacetime locality plays a crucial role in the appearance 
of an approximately classical world.  (We discuss the example of spin 
measurements from this perspective in the appendix.)  We also argue 
that the dimension of the Hilbert space for the entire universe must 
be infinite in order to be consistent with the fact that we observe an 
ordered world obeying the laws of physics.  This provides an important 
constraint on the structure of the fundamental theory.

In the second half of the paper, 
Sections~\ref{sec:locality-grav}~--~\ref{sec:measure}, we 
discuss quantum mechanics in a system with gravity.  This part is 
largely elucidation of the results obtained in Ref.~\cite{Nomura:2011dt} 
from a clearer perspective of giving quantum mechanical description 
of a general relativistic system.  There are, however, some important 
refinements, e.g.\ on the precise definition of ${\cal M}$ in 
Eq.~(\ref{eq:multiverse-H}), the treatment of spacetime singularities, 
and a useful probability formula applying in many practical cases. 
In Section~\ref{sec:locality-grav} we argue that spacetime locality, 
which plays a crucial role in understanding quantum measurement, 
is preserved only if we define quantum states in appropriately 
restricted spacetime regions.  In Section~\ref{sec:Hilbert-frame} 
we determine the Hilbert space of Eq.~(\ref{eq:multiverse-H}) 
from the viewpoint of fixing/eliminating all the redundancies 
and overcountings in a general relativistic, global spacetime 
description of nature.  We argue that complementarity as well 
as the observer dependence of horizons can be understood in a 
unified manner from the fact that changes of the reference frame 
are represented in the Hilbert space defined in restricted spacetime 
regions.  In Section~\ref{sec:Hilbert-QG} we reproduce the argument 
of Ref.~\cite{Nomura:2011dt}, which says that the multiverse state evolves 
asymptotically into a supersymmetric Minkowski world, with an important 
modification associated with the treatment of spacetime singularities. 
This determines the complete Hilbert space structure for quantum 
gravity, and provides the required infinite dimensionality to explain 
our ordered observation.  In Section~\ref{sec:measure} we discuss 
probabilities in the quantum mechanical multiverse.  We restate that 
the eternally inflating multiverse and many worlds in quantum mechanics 
are the same.  Finally, in Section~\ref{sec:summary} we briefly 
summarize the whole paper.

Relations between quantum mechanics and the multiverse have been 
discussed in other work as well.  Reference~\cite{Bousso:2011up} 
considered the issue of basis selection in the context of the multiverse, 
although the resulting picture is crucially different from the one 
here, especially about unitarity of quantum mechanics.  Earlier 
considerations of quantum mechanics in the multiverse/universe can be 
found in Ref.~\cite{Aguirre:2010rw}.  Reference~\cite{Freivogel:2006xu} 
uses supersymmetric Minkowski space as an important ingredient of their 
proposal, although in a different way than in ours.  The picture of the 
multiverse from a local viewpoint, which arises here as a consequence 
of quantum mechanics, has been promoted in the context of geometric 
cutoff measures; see Ref.~\cite{Bousso:2006ev} for example.

\section{Probabilistic Interpretation of Quantum Mechanics}
\label{sec:interpret}

In this and the next two sections, we discuss how the probabilistic 
interpretation of quantum mechanics arises in a {\it complete} quantum 
mechanical system that {\it includes a physical observer} who actually 
measures physical quantities.  Note that in order to provide a full 
account of the measurement process, both the observer and an experimental 
apparatus must be a part of the description.  Having a precise understanding 
of this process is crucial to apply quantum mechanics to the entire 
universe (or the eternally inflating multiverse).

\subsection{Quantum measurement a la von~Neumann}
\label{subsec:modern}

Let us begin our discussion with a simple nonrelativistic state in two 
dimensional Hilbert space, ignoring the experimental apparatus and observer 
for the moment.  For definiteness, we take this to be a spin-$1/2$ system:
\begin{equation}
  \ket{\Psi_{\rm sys}} = c_\uparrow \ket{\uparrow} 
    + c_\downarrow \ket{\downarrow},
\label{eq:NR-1}
\end{equation}
where $|c_\uparrow|^2 + |c_\downarrow|^2 = 1$.  The conventional Copenhagen 
interpretation says that if we measure the spin of this system at some 
time $t = t_{\rm m}$, then we find it up or down with the probabilities 
$P_{\uparrow} = |c_\uparrow|^2$ and $P_{\downarrow} = |c_\downarrow|^2$, 
respectively.  We may write this as
\begin{equation}
  P_\alpha = \bra{\Psi_{\rm sys}} {\cal O}_{{\rm spin},\alpha}
    \ket{\Psi_{\rm sys}}
\qquad
  (\alpha = \uparrow,\downarrow),
\label{eq:NR-P_alpha}
\end{equation}
where ${\cal O}_{{\rm spin},\alpha}$ is the operator that acts on 
$\ket{\Psi_{\rm sys}}$ and projects onto the state with a definite spin 
$\alpha$, e.g.\ ${\cal O}_{{\rm spin},\uparrow} \ket{\uparrow} = 1$ 
and ${\cal O}_{{\rm spin},\uparrow} \ket{\downarrow} = 0$.%
\footnote{For simplicity, here we focus only on a single direction of 
 the spin.  Including other directions does not affect the argument below.}
After we measure a definite outcome, e.g.\ spin up, the wavefunction 
$\ket{\Psi_{\rm sys}}$ of the system ``collapses''
\begin{equation}
  \ket{\Psi_{\rm sys}} \stackrel{t=t_{\rm m}}{\longrightarrow} 
    \ket{\uparrow},
\label{eq:NR-collapse}
\end{equation}
so that subsequent measurements will always find the spin pointing up.

In a modern viewpoint, a physical measurement is treated as interactions 
between the measured system (the spin-$1/2$ system under the case 
considered) and an experimental apparatus, as discussed originally 
by von~Neumann~\cite{von_Neumann}.  The process described above then 
corresponds to the following situation.  For $t \ll t_{\rm m}$, the 
combined state of the spin and the apparatus is
\begin{equation}
  \ket{\Psi_{{\rm sys}+{\rm app}}(t \ll t_{\rm m})} 
  = \bigl( c_\uparrow \ket{\uparrow} + c_\downarrow \ket{\downarrow} \bigr) 
    \otimes \bigl|\apparatusneut\bigr>,
\label{eq:NR-before}
\end{equation}
where $\bigl|\apparatusneut\bigr>$ represents the apparatus being 
in a ``ready'' state.  (We take the Schr\"{o}dinger picture throughout.) 
After $t \approx t_{\rm m}$, the full state becomes
\begin{equation}
  \ket{\Psi_{{\rm sys}+{\rm app}}(t \gg t_{\rm m})} = 
    c_\uparrow \ket{\uparrow} \otimes \bigl|\apparatusup\bigr> 
    + c_\downarrow \ket{\downarrow} \otimes \bigl|\apparatusdown\bigr>,
\label{eq:NR-after}
\end{equation}
due to the {\it standard time evolution} of the state:\ 
$\ket{\Psi_{{\rm sys}+{\rm app}}(t_1)} = e^{-iH(t_1-t_2)} 
\ket{\Psi_{{\rm sys}+{\rm app}}(t_2)}$.  Here, $\bigl|\apparatusup\bigr>$ 
and $\bigl|\apparatusdown\bigr>$ represent configurations of the 
apparatus showing that it has measured spin up and down, respectively, 
and $H$ is the Hamiltonian for the combined system of the apparatus 
and spin.  This particular process, making the state of the apparatus 
entangled with that of the measured system, is called decoherence in the 
narrow sense~\cite{Schlosshauer}.  A striking fact is that the dynamical 
evolution from Eq.~(\ref{eq:NR-before}) to Eq.~(\ref{eq:NR-after}) 
occurs very quickly; namely, a microscopic system can affect a macroscopic 
system drastically in a rather short timescale, in a way that they can 
no longer be considered independent, separate systems.  This is a crucial 
aspect of quantum mechanics that makes it hard to grasp using classical 
intuition.

\subsection{The problem of the preferred basis}
\label{subsec:preferred-basis}

It is tempting to interpret Eq.~(\ref{eq:NR-after}) to show that 
the apparatus always measures either spin up or down, and not 
their superpositions.%
\footnote{Of course, this statement applies only if Eq.~(\ref{eq:NR-before}) 
 becomes Eq.~(\ref{eq:NR-after}) under deterministic, quantum mechanical 
 evolution.  If the experimental apparatus is designed such that it can 
 measure interference effects, then the state after the measurement does 
 not take the simple form of Eq.~(\ref{eq:NR-after}).}
In fact, if the measured system is not a single spin, but a macroscopic 
object such as a chair, then the above discussion seems to explain the 
fact that we never observe superpositions of a macroscopic object in our 
everyday experience.  Equation~(\ref{eq:NR-after}) alone, however, is 
not enough to show this because it can also be written in an arbitrary 
basis as~\cite{Zurek:1981xq}
\begin{equation}
  \ket{\Psi_{{\rm sys}+{\rm app}}(t \gg t_{\rm m})} = 
    \ket{1} \otimes \bigl|\apparatusone\bigr> 
    + \ket{2} \otimes \bigl|\apparatustwo\bigr>.
\label{eq:NR-basis}
\end{equation}
Here,
\begin{equation}
  \bigl|\apparatusi\bigr> = \sum_{\alpha=\uparrow,\downarrow} 
    U_{i\alpha} \bigl|\apparatusalpha\bigr>
\qquad
  (i = 1,2)
\label{eq:app-basis}
\end{equation}
is a basis for the apparatus, and
\begin{equation}
  \ket{i} = \sum_{\alpha=\uparrow,\downarrow} 
    c_\alpha\, (U^{-1})_{\alpha i} \ket{\alpha}
\label{eq:sys-basis}
\end{equation}
the corresponding states for the measured system, where $U$ is an 
arbitrary $2 \times 2$ unitary matrix.  In particular, if $|c_\uparrow| 
= |c_\downarrow|$, then the states $\ket{1}$ and $\ket{2}$ form an 
orthogonal basis, so that they can be eigenstates of some Hermitian 
operator.  How can one then say that the apparatus has measured the 
system in the $\{ \ket{\uparrow}, \ket{\downarrow} \}$ basis, not 
in the $\{ \ket{1}, \ket{2} \}$ basis?

The ambiguity of the basis described above has been confusing some 
fundamental physicists.  A standard answer to this question is 
{\it environmental decoherence}~\cite{Zurek:1981xq,Zurek:1982ii}, whose 
implementation in the present context goes as follows.  We first regard 
the apparatus and spin as {\it open} quantum systems, interacting with 
some ``environment'' $\ket{E_0}$.  We can then define {\it the preferred 
states} for the combined apparatus-spin system as the states that are 
least sensitive to the interaction with the environment, i.e.\ those 
that are least entangled with the environment by dynamical evolution. 
For instance, if the interaction between the apparatus and environment 
is such that
\begin{equation}
\left\{ \begin{array}{l}
  \bigl|\apparatusup\bigr> \otimes \bigl| E_0 \bigr> \rightarrow 
  \bigl|\apparatusup\bigr> \otimes \bigl| E_1 \bigr> \\
  \bigl|\apparatusdown\bigr> \otimes \bigl| E_0 \bigr> \rightarrow
  \bigl|\apparatusdown\bigr> \otimes \bigl| E_2 \bigr>,
\end{array} \right.
\label{eq:app-env}
\end{equation}
with $\left< E_1 | E_2 \right> \rightarrow 0$, then the preferred states 
are the two terms in the right-hand side of Eq.~(\ref{eq:NR-after}) because 
each of them will {\it not} get entangled with the environment according 
to Eq.~(\ref{eq:app-env}).  (Here, we have ignored the interaction 
between the spin and environment.)  The measurement is then claimed 
to be performed in this preferred state basis.

In this picture of environment-induced basis selection, the openness 
of quantum systems plays a crucial role in understanding measurement 
processes.  In fact, such a picture is appropriate for the purpose of 
discussing consequences of quantum measurement performed in terrestrial 
experiments, which are indeed open.  At the fundamental level, however, 
this raises the following question:\ what if we include the environment 
in the description of our quantum state?  One might say that there 
is always some environment for any system {\it in practice}, but here 
we are talking about the fundamental issue.  This question becomes 
particularly acute if we try to apply quantum mechanics to describe 
the entire universe, since then it is not even clear what one can take 
as an environment {\it for the entire universe}.

A line of reasoning like this has recently led the authors of 
Ref.~\cite{Bousso:2011up} to claim that quantum mechanics is operationally 
well defined {\it only} under the existence of {\it intrinsically 
inaccessible} degrees of freedom, which they took to be those escaping 
a cosmic horizon in the eternally inflating multiverse.  In this picture, 
quantum mechanical evolution is intrinsically irreversible---to obtain 
probabilistic interpretation of quantum mechanics, degrees of freedom 
outside the horizon {\it must} be traced out.  Here we will argue 
differently---we need not introduce such irreversibility at the fundamental 
level.  We argue that, as discussed in Ref.~\cite{Nomura:2011dt}, the 
principles of quantum mechanics, including deterministic {\it unitary} 
evolution of the states, are fully respected if one describes physics 
as viewed from a single reference frame.%
\footnote{In Ref.~\cite{Nomura:2011dt}, the word ``observer'' was used 
 to denote a reference frame in which the quantum multiverse is described. 
 The issue of the reference frame in a quantum mechanical system with 
 gravity will be discussed in detail in Section~\ref{sec:Hilbert-frame}, 
 where we will see that quantum mechanics forces us to describe the 
 system from the viewpoint of a single ``observer.''}
The effective irreversibility of the quantum-to-classical transition 
appears simply because the dimension of Hilbert space is infinite.  The 
ambiguity of the basis is fixed by a feature in the dynamics, specifically 
spacetime locality as encoded in the algebra of (low energy) operators. 
We now see in detail what the implications of spacetime locality 
are in our context.

\section{Physical Predictions and Spacetime Locality}
\label{sec:locality}

To follow what a measurement {\it actually} means in a realistic context, 
here we consider a more elaborate model of the setup discussed in the 
previous section.  We will see that all the physical information is 
encoded in {\it matrix elements}, and that the ultimate answer to the 
basis question should lie in the algebra of quantum operators and the 
dynamics associated with it.

\subsection{Physical information is in matrix elements}
\label{subsec:relative}

Suppose that at early times $t \ll t_{\rm m}$, the detector apparatus 
has not yet interacted with the spin, as in Eq.~(\ref{eq:NR-before}). 
We assume that the detector is located on a desk, which we also include 
in our description.  Moreover, we also consider an observer who has not 
initially been looking at the detector.  The initial state of this entire 
system is then
\begin{equation}
  \ket{\Psi(t \ll t_{\rm m})} 
  = \bigl( c_\uparrow \ket{\uparrow} + c_\downarrow \ket{\downarrow} \bigr) 
    \otimes \bigl|\apparatusneut\bigr> \otimes \bigl|\desk\bigr> 
    \otimes \bigl|\obsbefore\bigr>.
\label{eq:Psi-1}
\end{equation}
Here, we do not necessarily consider that $c_\uparrow$ and $c_\downarrow$ 
are normalized as $|c_\uparrow|^2 + |c_\downarrow|^2 = 1$ (anticipating 
that the state in general may have terms additional to the ones shown 
above).  At $t \approx t_{\rm m}$, the detector interacts with the 
spin, but we assume that the observer does not look at it until a later 
time $t_{\rm obs} > t_{\rm m}$, so
\begin{equation}
  \ket{\Psi(t_{\rm m} \ll t \ll t_{\rm obs})} 
  = \Bigl( c_\uparrow \ket{\uparrow} \otimes \bigl|\apparatusup\bigr> 
    + c_\downarrow \ket{\downarrow} \otimes \bigl|\apparatusdown\bigr> \Bigr) 
    \otimes \bigl|\desk\bigr> \otimes \bigl|\obsbefore\bigr>.
\label{eq:Psi-2}
\end{equation}
Finally, at $t = t_{\rm obs}$, the observer reads what the detector shows, 
and his/her brain state reacts accordingly:
\begin{equation}
  \ket{\Psi(t \gg t_{\rm obs})} 
  = c_\uparrow \ket{\uparrow} \otimes \bigl|\apparatusup\bigr> 
    \otimes \bigl|\desk\bigr> \otimes \bigl|\obsafter\brainup\bigr> 
    + c_\downarrow \ket{\downarrow} \otimes \bigl|\apparatusdown\bigr> 
    \otimes \bigl|\desk\bigr> \otimes \bigl|\obsafter\braindown\bigr>.
\label{eq:Psi-3}
\end{equation}
Note that the time evolution of the state, Eq.~(\ref{eq:Psi-1}) 
$\rightarrow$ Eq.~(\ref{eq:Psi-2}) $\rightarrow$ Eq.~(\ref{eq:Psi-3}), 
is caused by the standard, deterministic quantum evolution:\ 
$\ket{\Psi(t_1)} = e^{-iH(t_1-t_2)} \ket{\Psi(t_2)}$, where $H$ 
is the Hamiltonian for the combined system of the spin, apparatus, 
desk, and the observer.

According to the standard rule of quantum mechanics, we expect that 
the probabilities for the observer to measure spin up and down should 
respectively be
\begin{equation}
  P_\alpha = \frac{|c_\alpha|^2}{|c_\uparrow|^2 + |c_\downarrow|^2}
\qquad
  (\alpha = \uparrow,\downarrow),
\label{eq:NR-P_alpha-2}
\end{equation}
ignoring the issue of the basis ambiguity.  What is the {\it precise} 
meaning of this equation?  Note that the question we are asking here 
is actually the following:\ assuming that the observer learns the result 
of the experiment by reading the apparatus, what does he/she find? 
This conditional probability is given, according to the standard 
Born rule, by
\begin{equation}
  P(\alpha|{\rm obs}) = 
    \frac{\bra{\Psi(t \gg t_{\rm obs})} {\cal O}_{{\rm obs},\alpha} 
      \ket{\Psi(t \gg t_{\rm obs})}}
    {\bra{\Psi(t \gg t_{\rm obs})} {\cal O}_{\rm obs} 
      \ket{\Psi(t \gg t_{\rm obs})}}
\qquad
  (\alpha = \uparrow,\downarrow),
\label{eq:NR-P_alpha-3}
\end{equation}
where ${\cal O}_{{\rm obs},\alpha}$ is the operator projecting onto the 
state in which the observer learns the result to be $\alpha$
\begin{equation}
  {\cal O}_{{\rm obs},\alpha} 
  = {\bf 1} \otimes {\bf 1} \otimes {\bf 1} \otimes \Bigl\{ 
    \bigl|\obsafter\brainalpha\bigr> \bigl<\obsafter\brainalpha\bigr| \Bigr\}
\label{eq:O_obs-alpha}
\end{equation}
and ${\cal O}_{\rm obs}$ onto the one in which he/she learns {\it some} 
result, whatever it is:
\begin{equation}
  {\cal O}_{\rm obs} = \sum_\alpha {\cal O}_{{\rm obs},\alpha}.
\label{eq:O_obs-any}
\end{equation}
Here, the states appearing in Eq.~(\ref{eq:NR-P_alpha-3}) have been 
chosen simply as $\ket{\Psi(t \gg t_{\rm obs})}$, since $\ket{\Psi(t)}$ 
is assumed to be (approximately) time independent for $t \gg t_{\rm obs}$. 
The case in which a state has general time dependence, leading 
to Eq.~(\ref{eq:probability-AB}) (or its generalization, 
Eq.~(\ref{eq:prob-final})), will be discussed carefully later 
(in Section~\ref{sec:measure}).

The simple analysis above highlights some of the important features of 
our present description of quantum measurement.  First, expressed in 
the form of conditional probabilities as in Eq.~(\ref{eq:NR-P_alpha-3}), 
physical predictions do {\it not} depend on how the state is written, 
{\it including in what basis it is expanded}.  This property is obvious 
in Eq.~(\ref{eq:NR-P_alpha-3}), since the matrix elements do not 
depend on the basis used to expand the state, but it is obscured if 
one focuses only on the state and its expansion coefficients, as in 
Eq.~(\ref{eq:NR-P_alpha-2}).  Second, our (more fundamental) formula 
of Eq.~(\ref{eq:NR-P_alpha-3}) reproduces Eq.~(\ref{eq:NR-P_alpha-2}) 
{\it even if the state $\ket{\Psi(t)}$ contains additional terms that 
are not selected by the projection operator ${\cal O}_{\rm obs}$}. 
In fact, given an initial condition, the state at late times might 
contain a term representing a possibility that is not listed in 
Eq.~(\ref{eq:Psi-3}); for example, the apparatus might break before 
the observer reads it, or the observer might change his/her mind and 
never look at the apparatus.  (Note that, once the initial condition 
is given, the future state $\ket{\Psi(t)}$ is uniquely determined 
according to the deterministic quantum evolution specified by $H$, 
i.e., the choice is not left to us to eliminate these ``unwanted'' 
possibilities.)  Because of the way we asked the question, however, 
our answer always satisfies
\begin{equation}
  \sum_\alpha P(\alpha|{\rm obs}) = 1.
\label{eq:P-sum}
\end{equation}
Namely, the possible additional terms in the state $\ket{\Psi(t)}$ (or 
additional ``components'' in the wavefunction) are irrelevant for the 
question we are asking.

As discussed in detail in Ref.~\cite{Nomura:2011dt}, {\it any} physical 
question can be phrased in the form of a conditional probability; 
in the simplest setup, we can phrase it as:\ ``Given what we know 
about our past light cone, $A$, what is the probability of that 
light cone to have properties $B$ as well?''  This eliminates the 
question of ``What is the right basis {\it to expand the state}?'' 
The answer is that ``It doesn't matter.''  Once the question is 
phrased in this way using the appropriate projection operators 
${\cal O}_A$ and ${\cal O}_{A \cap B}$ (e.g.\ ${\cal O}_{\rm obs}$ 
and ${\cal O}_{{\rm obs},\alpha}$ in the above example), the desired 
probability $P(B|A)$ is defined unambiguously.  Is this sufficient 
to eliminate the basis ambiguity for quantum measurement, discussed 
in Section~\ref{subsec:preferred-basis}?  The answer is no---we 
need to discuss quantum operators, especially ${\cal O}_A$ and 
${\cal O}_{A \cap B}$, to see if there is any ambiguity there.

\subsection{Spacetime locality and the basis in Hilbert space}
\label{subsec:locality-basis}

We have seen that once a physical question is phrased in terms of 
${\cal O}_A$ and ${\cal O}_{A \cap B}$, the answer is unambiguously given 
by the probability formula, such as Eq.~(\ref{eq:NR-P_alpha-3}).  But, 
how can these operators be constructed?  In particular, is there any 
ambiguity in writing these operators (and if so, wouldn't that just 
be trading the basis ambiguity of states for that of operators)?

First of all, we note that it is appropriate to discuss the issue of 
basis in terms of operators, rather than states, as we will do here. 
This is because Hilbert space by itself does not carry any physical 
information other than its dimensionality---any (complex) Hilbert spaces 
having the same dimension are identical with each other.  {\it All the 
information about a physical system (except for its dimensionality) 
is encoded in quantum operators and the algebra they satisfy.}  Of 
course, being operators acting on a vector space, these quantum operators 
may also be written in an arbitrary basis.  However, we now have 
dynamical structures that may distinguish some basis over the others. 
In particular, there can be a special basis in which algebraic relations 
among operators look particularly simple.

Consider a (special) relativistic system.  At length scales much larger 
than the possible quantum gravity scale (and the entropy density much 
lower than that of a black hole), such a system is described by quantum 
field theory~\cite{Weinberg:QFT}.  Suppose there is only a single species 
of particles, represented by a quantum field $\phi({\bf x})$ with ${\bf x}$ 
being spatial coordinates.  The field $\phi({\bf x})$ at different 
${\bf x}$ should be regarded as different operators, which satisfy
\begin{equation}
  [ \phi({\bf x}), \pi({\bf x'}) ]_\mp = i\, \delta_{{\bf x},{\bf x'}},
\qquad
  [ \phi({\bf x}), \phi({\bf x'}) ]_\mp 
  = [ \pi({\bf x}), \pi({\bf x'}) ]_\mp = 0,
\label{eq:QFT-algebra}
\end{equation}
where $\pi({\bf x})$ is a conjugate momentum of $\phi({\bf x})$, and 
$[\, ,\, ]_\mp$ represents a commutator and anti-commutator if the particle 
is a boson and fermion, respectively.  Here, we have discretized spatial 
coordinates ${\bf x}$ for presentation purposes.%
\footnote{Since we consider low energy physics, some sort of discretization, 
 e.g.\ a finite (effective) ultraviolet cutoff, is appropriate.  Note 
 that here we use the Schr\"{o}dinger picture, which is unconventional 
 in quantum field theory.}
An important point is that in this ``local field'' basis, the time evolution 
operator $U(t_1,t_2) = e^{-iH(t_1-t_2)}$ takes a particularly simple form
\begin{equation}
  H = \sum_{\bf x}
    H_{\bf x}(\phi({\bf x}+\boldsymbol\epsilon_0),
      \phi({\bf x}+\boldsymbol\epsilon_1),\cdots; 
    \pi({\bf x}+\boldsymbol\epsilon_0),
      \pi({\bf x}+\boldsymbol\epsilon_1),\cdots),
\label{eq:QFT-local}
\end{equation}
{\it where $\boldsymbol\epsilon_i$ ($i = 0,1,2,\cdots$) runs only over 
a very small subset of the coordinates around $\boldsymbol\epsilon_0 
\equiv {\bf 0}$ (typically ``nearest neighbors'')}.  This is not the case 
if we use an arbitrary basis
\begin{equation}
  \phi({\bf z}) = \sum_{\bf x} c_{{\bf z} {\bf x}} \phi({\bf x}) 
\quad
  (c_{{\bf z} {\bf x}} \neq \delta_{{\bf z} {\bf x}}).
\label{eq:QFT-another}
\end{equation}
Namely, if we write $H$ in terms of $\phi({\bf z})$ and $\pi({\bf z})$ 
in the form of Eq.~(\ref{eq:QFT-local}), then $\boldsymbol\epsilon_i$ 
need not run only over a very small subset of the coordinates for 
generic $c_{{\bf z} {\bf x}}$.

The {\it existence} of a special basis satisfying Eq.~(\ref{eq:QFT-local}) 
is exactly what we call spacetime locality.  This is a property of nature 
whose origin is not yet fully understood---it is simply an empirical 
fact that there is such a basis at length scales that have been probed 
experimentally so far.%
\footnote{It is important to realize that spacetime locality is the 
 property of the operator algebra, and not states.  In fact, a state can 
 be easily (and, indeed, is generically) non-local, e.g., as the Bell 
 state appearing in the Einstein-Podolsky-Rosen experiment.}
This property, however, is crucial in selecting a particular basis {\it 
in Hilbert space} in which a simple description of physics is obtained. 
Specifically, consider a set of (time-independent) states $\ket{\kappa_m}$ 
that are eigenstates of the particle-number operators $N_{\bf x}$ 
{\it for all ${\bf x}$} (not ${\bf z}$):
\begin{equation}
  N_{\bf x}(\phi({\bf x}),\pi({\bf x})) \ket{\kappa_m} 
  = n_{\bf x}^{(m)} \ket{\kappa_m}.
\label{eq:kappa_m}
\end{equation}
These states are special in that they have well-defined configurations 
in physical space ${\bf x}$.  Furthermore, since the set of states in 
Eq.~(\ref{eq:kappa_m}) spans Fock space, it can form an orthonormal 
basis of Hilbert space; namely, an arbitrary state $\ket{\Psi(t)}$ may 
be written as a superposition
\begin{equation}
  \ket{\Psi(t)} = \sum_m c_m(t) \ket{\kappa_m},
\label{eq:Psi-t_gen}
\end{equation}
where $c_m(t)$ are complex functions and $\inner{\kappa_m}{\kappa_n} 
= \delta_{mn}$.  Note that the particular basis here, 
$\ket{\kappa_m}$, has been chosen such that an algebraic relation 
between operators---specifically the form of $H$ in terms of 
$\phi({\bf x})$ and $\pi({\bf x})$---takes a simple form in that 
basis.  In fact, the very concept of ``configurations in space'' 
arises as a result of the special property in Eq.~(\ref{eq:QFT-local}); 
without that, ${\bf x}$ could not even be interpreted as spatial 
coordinates.

We should emphasize that {\it the choice of the Hilbert space 
basis} discussed here does not by itself address the issue of 
{\it basis selection for quantum measurement} described in 
Section~\ref{subsec:preferred-basis}, although the former is needed 
for the discussion of the latter.  Indeed, the choice described here 
is, in some sense, ``a matter of convenience,'' in that we can also 
describe physics using the $\phi({\bf z})$ basis in principle (because 
the matrix elements, appearing in the probability formula, do not 
depend on the basis).  In this basis, however, the time evolution 
operator has an extremely complicated form, which completely obscures 
the fact that the dynamics respects spacetime locality.%
\footnote{As an analogy, one can imagine describing high energy behaviors 
 of QCD in the gravitational picture, using the gauge/gravity duality. 
 In that picture, any high energy scattering will have contributions 
 from complicated high-curvature stringy effects, which completely 
 obscures the fact that it is given by a simple, perturbative 
 gluon exchange amplitude.}
Therefore, in practice one always needs to choose a Hilbert space basis 
associated with locality:\ either $\ket{\kappa_m}$ in Eq.~(\ref{eq:kappa_m}) 
or a basis that has a simple relation to it (such as the momentum basis).

\subsection{What physical questions may one ask?}
\label{subsec:questions}

Let us choose the ``locality basis'' $\ket{\kappa_m}$, given in 
Eq.~(\ref{eq:kappa_m}).  Then there is no ambiguity in expanding states 
as in Eq.~(\ref{eq:Psi-t_gen}).  The question, however, still remains:\ 
how can we choose the ``correct form'' for projection operators ${\cal O}_A$ 
and ${\cal O}_{A \cap B}$ appearing in the probability formula?  Experience 
says that all the information we can explicitly handle (in the sense 
that it can be duplicated in physical systems) is given in the form 
of, e.g., Eqs.~(\ref{eq:NR-P_alpha-3}~--~\ref{eq:O_obs-any})---i.e.\ by 
operators projecting onto states that have {\it well-defined macroscopic 
configurations in the phase space (up to some uncertainties)}.  Why 
is that?

In general, states having well-defined macroscopic configurations, e.g.\ 
$\bigl|\apparatusalpha\bigr>$, $\bigl|\desk\bigr>$, $\bigl|\obsbefore\bigr>$, 
and $\bigl|\obsafter\brainalpha\bigr>$ in previous sections, are obtained 
as superpositions of $\ket{\kappa_m}$ that have ``similar'' spatial 
configurations.  For each macroscopic configuration $i$, we have a set 
of $n_i$ corresponding microstates
\begin{equation}
  \ket{\psi^{\rm (i)}_a} = \sum_m f_{a,m}^{(i)} \ket{\kappa_m}
  \quad (a = 1,\cdots,n_i),
\label{eq:psi-ia}
\end{equation}
which we collectively call $\ket{\alpha_i}$.  Here, $f_{a,m}^{(i)}$ for 
each $(i,a)$ play the role of a smearing function in position space, 
ensuring that the configuration has a well-defined momentum at the 
macroscopic level.  The projection operator onto macroscopic configuration 
$i$ can then be defined as
\begin{equation}
  \ket{\alpha_i} \bra{\alpha_i} \equiv 
    \sum_{a=1}^{n_i} \ket{\psi^{\rm (i)}_a} \bra{\psi^{\rm (i)}_a},
\label{eq:project-i}
\end{equation}
where we have taken $\bigl< \psi^{\rm (i)}_a | \psi^{\rm (i)}_b \bigr> 
= \delta_{ab}$.  Since $\bigl< \psi^{\rm (i)}_a | \psi^{\rm (j)}_b \bigr> 
\approx 0$ for different macroscopic configurations $i \neq j$, these 
projection operators satisfy $P_i P_j \approx P_j P_i \approx 0$ for 
$i \neq j$, where $P_i = \ket{\alpha_i} \bra{\alpha_i}$.

The issue is why physical questions we ask are always phrased in 
terms of ${\cal O}_A$ and ${\cal O}_{A \cap B}$ that take the form
\begin{equation}
  {\cal O}_A = \sum_{i \in A} \ket{\alpha_i} \bra{\alpha_i}
\label{eq:O_A-local}
\end{equation}
(and similarly for ${\cal O}_{A \cap B}$), where $i \in A$ implies 
that the sum is taken for the configurations that satisfy condition $A$. 
In particular, what is wrong with using $\ket{\alpha_i}$ corresponding 
to a superposition of macroscopically different configurations, i.e.\ 
microstates $\bigl| \psi^{\rm (i)}_a \bigr>$ in which the expansion 
coefficients $f_{a,m}^{(i)}$ have significant supports from {\it 
macroscopically} different configurations $m$?  This strong restriction 
on possible questions we may ask is the essence of basis selection for 
quantum measurement, and composes what we call the quantum-to-classical 
transition.  Its origin is in the dynamics, specifically spacetime 
locality as encoded in Eq.~(\ref{eq:QFT-local}), as we will discuss 
in the next section.

\section{Classical Reality in Quantum Mechanical Systems}
\label{sec:QM-classical}

In this section, we discuss the origin of the following basic observational 
fact:\ we perceive our macroscopic world to be classical, having a 
well-defined configuration in phase space.  In fact, this statement 
consists of at least two elements, which are respectively to do with 
``basis selection'' and ``wavefunction collapse'' of quantum measurement:
\begin{itemize}
\item[(i)] Probabilistic processes in quantum mechanics are well described 
by density matrices that are diagonal in the ``classical state basis'' 
$\ket{\alpha_i}$, at least for macroscopic systems.
\item[(ii)] A measurement selects an outcome; namely, we can ignore other 
possible outcomes after a measurement is actually performed.
\end{itemize}
In the standard treatment of these problems, the openness of a system 
is emphasized~\cite{Schlosshauer}.  Here we ask if the openness is really 
necessary at the fundamental level to account for these features.  We 
will argue that the answer is no---the quantum-to-classical transition 
may occur consistently with observation even in a closed quantum 
mechanical system.

On the other hand, we will also argue that cosmology based on a closed 
system fails to explain another basic element necessary to explain the 
observational fact:
\begin{itemize}
\item[(iii)] We observe an ordered world, i.e., we {\it perceive} a world 
that obeys consistent laws of physics.
\end{itemize}
This argument will force us to consider that the Hilbert space for the 
entire universe (multiverse) is infinitely large:\ ${\rm dim}\,{\cal H} 
= \infty$, unless we abandon unitarity of quantum mechanical evolution. 
In the actual universe, this requirement is satisfied because it 
asymptotically becomes a supersymmetric Minkowski (or singularity) 
world, which we will see in more detail in Section~\ref{sec:Hilbert-QG}.

\subsection{A double-slit experiment in a large system}
\label{subsec:double-slit}

We begin by a standard analysis of the double-slit experiment, which 
sets the stage for later discussions.  For the moment, we can be agnostic 
about whether the entire system is open or closed.  Our analysis applies 
as long as the dynamical timescale is much shorter than the thermalization 
timescale $t_{\rm th}$, which is indeed the case for a sufficiently large 
system ($t_{\rm th} = \infty$ for an open system).

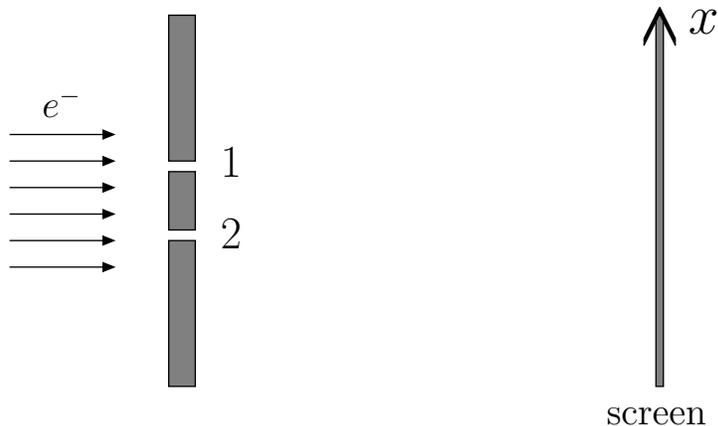
\begin{figure}[t]
\begin{center}
\begin{picture}(240,175)(80,-20)
  \Text(75,102)[b]{\large $e^-$}
  \LongArrow(55,95)(95,95) \LongArrow(55,85)(95,85)
  \LongArrow(55,75)(95,75) \LongArrow(55,65)(95,65)
  \LongArrow(55,55)(95,55) \LongArrow(55,45)(95,45)
  \Text(135,85)[l]{\Large $1$} \Text(135,58)[l]{\Large $2$}
  \GBox(115,0)(125,55){0.5} \GBox(115,59)(125,81){0.5}
  \GBox(115,85)(125,140){0.5}
  \GBox(298.6,0)(301.4,140){0.5}  \Text(312,138)[l]{\LARGE $x$}
  \Line(300,143)(294,131) \Line(300,143)(306,131)
  \Line(300,142.5)(294,130.5) \Line(300,142.5)(306,130.5)
  \Line(300,142)(294,130) \Line(300,142)(306,130)
  \Line(300,141.5)(294,129.5) \Line(300,141.5)(306,129.5)
  \Line(300,141)(294,129) \Line(300,141)(306,129)
  \Line(300,140.5)(294,128.5) \Line(300,140.5)(306,128.5)
  \Text(300,-8)[t]{\large screen}
\end{picture}
\caption{The double-slit experiment with an electron.  The electron at 
 the two slits are represented by $\ket{1}$ and $\ket{2}$, respectively, 
 while that at the screen at position $x$ by $\ket{x}$.}
\label{fig:double-slit}
\end{center}
\end{figure}

The setup of the experiment is such that an electron, initially prepared 
at slits as
\begin{equation}
  \ket{\psi_{e,{\rm init}}} = \frac{1}{\sqrt{2}} (\ket{1} + \ket{2})
\label{eq:wf-slit}
\end{equation}
evolves according to
\begin{equation}
  \ket{1} \rightarrow \int\!\!dx\,\, \psi_1(x) \ket{x},
\qquad
  \ket{2} \rightarrow \int\!\!dx\,\, \psi_2(x) \ket{x}.
\label{eq:12-evolve}
\end{equation}
Here, $\ket{1}$ and $\ket{2}$ represent the electron localized at 
slits~1 and 2, respectively, while $\ket{x}$ represents the electron 
localized at position $x$ on the screen (see Fig.~\ref{fig:double-slit}). 
We consider sending only a single electron for the sake of simplicity.

The entire system consists of the electron, detector apparatus, and the 
rest of the world, which is initially in the state
\begin{equation}
  \ket{\Psi_{\rm init}} = 
    \biggl\{ \frac{1}{\sqrt{2}} (\ket{1} + \ket{2}) \biggr\}
    \otimes \bigl|\apparatusneut\bigr> 
    \otimes \bigl|\mbox{\large $R_{}$}\bigr>,
\label{eq:exp-init}
\end{equation}
where $\bigl|\apparatusneut\bigr>$ represents the detector in a ready 
state and $\bigl|\mbox{\large $R_{}$}\bigr>$ the degrees of freedom that 
are not included in the electron or detector state.%
\footnote{Here we consider $\bigl|\ftapparatusneut\bigr>$ and 
 $\bigl|\mbox{\normalsize $R_{}$}\bigr>$ to be one of the microstates 
 having these macroscopic configurations.  The projection operators 
 corresponding to macroscopic configurations, e.g.\ the one in 
 Eq.~(\ref{eq:O_x}) below, should then be interpreted as defined in 
 Section~\ref{subsec:questions}.  This is appropriate since we are 
 interested in the fundamental theory here.  In practice, we do not 
 know which microstate a given macroscopic configuration is in, so we 
 may consider a density matrix in which all the microscopic information 
 is coarse-grained.  Our formalism is straightforwardly extended 
 to this case; see e.g.\ Eq.~(\ref{eq:prob-final-rho}) in 
 Section~\ref{subsec:Born}.}
Assuming that there is no interaction between the experimental apparatus 
and the rest of the world, this state evolves into
\begin{equation}
  \ket{\Psi_{\rm fin}} = \frac{1}{\sqrt{2}} 
    \biggl( \int\!\!dx\,\,\psi_1(x) \ket{0} \otimes \bigl|\apparatusx\bigr> 
      + \int\!\!dx\,\,\psi_2(x) \ket{0} \otimes \bigl|\apparatusx\bigr> 
    \biggr) \otimes \bigl|\mbox{\large $R_{}$}\bigr>.
\label{eq:exp-fin}
\end{equation}
Here, we have assumed that the combined electron and detector system 
evolves as
\begin{equation}
  \ket{x} \otimes \bigl|\apparatusneut\bigr> \rightarrow 
    \ket{0} \otimes \bigl|\apparatusx\bigr>,
\label{eq:absord-e}
\end{equation}
where $\ket{0}$ implies that the electron has absorbed into the apparatus, 
and $\bigl|\apparatusx\bigr>$ represents the status of the detector showing 
that the electron has arrived at $x$ on the screen.  The probability density 
of finding the electron at $x$ in this experiment is then given by
\begin{equation}
  P(x|{\rm exp}) = \frac{\bra{\Psi_{\rm fin}} {\cal O}_x \ket{\Psi_{\rm fin}}}
    {\bra{\Psi_{\rm fin}} \int\! {\cal O}_x dx \ket{\Psi_{\rm fin}}},
\label{eq:DS-prob}
\end{equation}
where
\begin{equation}
  {\cal O}_x = {\bf 1} \otimes 
    \Bigl\{ \bigl|\apparatusx\bigr>\bigl<\apparatusx\bigr| \Bigr\} 
    \otimes {\bf 1}
\label{eq:O_x}
\end{equation}
is the projection operator extracting the result of the experiment. 
Plugging Eq.~(\ref{eq:exp-fin}) into Eq.~(\ref{eq:DS-prob}), we obtain 
the standard result:
\begin{equation}
  P(x|{\rm exp}) \propto |\psi_1(x)|^2 + |\psi_2(x)|^2 
    + 2\, {\rm Re}\{ \psi_1(x) \psi_2^*(x) \}.
\label{eq:result}
\end{equation}
Here, we have used $\bigr<\apparatusx \big| \apparatusy\bigr> = 
\delta(x-y)$, and the last term represents the interference effect.

Let us now consider the situation that the path of the electron is 
``monitored'' by interactions of the electron with other degrees 
of freedom, e.g.\ a detector located close to the slits.  Including 
these degrees of freedom in $\bigl|\mbox{\large $R_{}$}\bigr>$, the 
final state of the evolution is now given by
\begin{equation}
  \ket{\Psi_{\rm fin}} = \frac{1}{\sqrt{2}} 
    \biggl( \int\!\!dx\,\,\psi_1(x) \ket{0} \otimes \bigl|\apparatusx\bigr> 
      \otimes \bigl|\mbox{\large $R_1$}\bigr> 
    + \int\!\!dx\,\,\psi_2(x) \ket{0} \otimes \bigl|\apparatusx\bigr> 
      \otimes \bigl|\mbox{\large $R_2$}\bigr> \biggr),
\label{eq:exp-fin-dec}
\end{equation}
instead of Eq.~(\ref{eq:exp-fin}).  Here, $\bigl|\mbox{\large $R_1$}\bigr>$ 
and $\bigl|\mbox{\large $R_2$}\bigr>$ represent the status of the degrees 
of freedom after the electron passes slits $1$ and $2$, respectively. 
Applying Eq.~(\ref{eq:DS-prob}), the probability density of finding the 
electron at $x$ is now
\begin{equation}
  P(x|{\rm exp}) \propto |\psi_1(x)|^2 + |\psi_2(x)|^2 
    + 2\, {\rm Re}\bigl\{ \psi_1(x) \psi_2^*(x) 
      \bigl<\mbox{\large $R_2$} \big| \mbox{\large $R_1$}\bigr> \bigr\}.
\label{eq:result-dec}
\end{equation}
Therefore, once the interactions of the electron at the slits lead to 
distinct configurations for the surroundings depending on which slit it 
passes, i.e.\ $\bigl<\mbox{\large $R_2$} \big| \mbox{\large $R_1$}\bigr> 
\ll 1$, the interference term disappears.  This is the well-known result 
obtained by Wootters and Zurek in Ref.~\cite{Wootters:1979zz}.

The question, again, is why the result of the experiment is described 
using the operator in Eq.~(\ref{eq:O_x}).  In other words, why do we 
perceive the world in such a way that a macroscopic quantum system 
decoheres in the classical state basis---in this case, the location 
of the pointer of the detector apparatus?  In the standard explanation 
due to environment-induced basis selection, the openness of a system 
plays a crucial role.  Below, we will obtain (essentially) the same 
result without invoking an openness of the whole system, which elucidates 
the real origin of the basis selection in quantum measurement.

\subsection{Dynamical selection of a measurement basis---spacetime locality}
\label{subsec:collapse}

Let us keep following the state of the system after the double-slit 
experiment was performed.  For simplicity, here we focus on the case 
where the electron path was not monitored, but the same argument 
applies to the monitored case as well.  Denoting the evolution after 
the double-slit measurement by
\begin{equation}
  \ket{\psi_x(t=0)} \equiv \ket{0} \otimes \bigl|\apparatusx\bigr> 
    \otimes \bigl|\mbox{\large $R_{}$}\bigr>
\quad
  \longrightarrow
\quad
  \ket{\psi_x(t)},
\label{eq:evol-after}
\end{equation}
the state of the entire system is given by
\begin{equation}
  \ket{\Psi(t)} = \int\!\!dx\, c_x \ket{\psi_x(t)},
\label{eq:Psi-after}
\end{equation}
where $t$ is the time passed after the measurement, and $c_x \equiv 
(\psi_1(x) + \psi_2(x))/(\int\!dy\,|\psi_1(y) + \psi_2(y)|^2)^{1/2}$.

Now, let us imagine that some other experiment is performed in this system 
at a late time $t_{\rm exp}$, which is still earlier than $t_{\rm th}$. 
This experiment will involve only a very small subset of the degrees of 
freedom in $\ket{\Psi(t)}$, as we consider that the entire system is very 
large.  We isolate these degrees of freedom by writing $\ket{\psi_x(t)} 
= \ket{\phi_x(t)} \otimes \ket{r_x(t)}$:
\begin{equation}
  \ket{\Psi(t)} = \int\!\!dx\, c_x \ket{\phi_x(t)} \otimes \ket{r_x(t)},
\label{eq:Psi-exp}
\end{equation}
where the first factor represents degrees of freedom associated with the 
experiment while the second the rest.  The outcome of this experiment 
can be calculated, using the probability formula based on matrix 
elements as
\begin{eqnarray}
  P(A|{\rm exp}) &=& \frac{\bra{\Psi(t_{\rm exp})} 
      {\cal O}_A \ket{\Psi(t_{\rm exp})}} 
    {\bra{\Psi(t_{\rm exp})} \sum_A\! {\cal O}_A \ket{\Psi(t_{\rm exp})}}
\nonumber\\
  &=& \frac{\int\!dxdy\, c_x^* c_y \inner{r_x(t_{\rm exp})}{r_y(t_{\rm exp})} 
      \left< \phi_x(t_{\rm exp}) | {\cal O}_A | \phi_y(t_{\rm exp}) \right>} 
    {\int\!dxdy\, c_x^* c_y \inner{r_x(t_{\rm exp})}{r_y(t_{\rm exp})} 
      \left< \phi_x(t_{\rm exp}) | \sum_A\! {\cal O}_A 
      | \phi_y(t_{\rm exp}) \right>},
\label{eq:prob-ex}
\end{eqnarray}
where ${\cal O}_A$ is the projection operator acting on $\ket{\phi_x(t)}$ 
selecting the situation where the experiment is performed with a definite 
outcome $A$.

The expression of Eq.~(\ref{eq:prob-ex}) contains terms representing 
interference between different outcomes {\it of the first, double-slit 
experiment}, i.e.\ $x$ and $y$ with $x \neq y$.  These terms, however, 
disappear if $\inner{r_x(t_{\rm exp})}{r_y(t_{\rm exp})} \ll 1$ for 
$x \neq y$, in which case
\begin{eqnarray}
  P(A|{\rm exp}) &=& 
    \frac{\int\!dx\, |c_x|^2 \left< \phi_x(t_{\rm exp}) | 
      {\cal O}_A | \phi_x(t_{\rm exp}) \right>} 
    {\int\!dx\, |c_x|^2 \left< \phi_x(t_{\rm exp}) | 
      \sum_A\! {\cal O}_A | \phi_x(t_{\rm exp}) \right>}
\nonumber\\
  &=& \int\!dx\, |c_x|^2 \left< \phi_x(t_{\rm exp}) | 
      {\cal O}_A | \phi_x(t_{\rm exp}) \right>.
\label{eq:prob-ex-2}
\end{eqnarray}
Here, we have assumed $\left< \phi_x(t_{\rm exp}) | \sum_A\! {\cal O}_A | 
\phi_x(t_{\rm exp}) \right> = 1$, i.e.\ the second experiment occurs no 
matter what the outcome of the double-slit experiment.  The probabilities 
for the outcomes of the two experiments now follow what we expect 
classically:
\begin{equation}
  P(A|{\rm exp}) = \int\!\!dx\, P(x \,|\, \mbox{double-slit exp})\, 
    P_x(A \,|\, \mbox{second exp}).
\label{eq:P-classical}
\end{equation}
Note that the result of the second experiment may depend on that of 
the first, as indicated by the subscript $x$ on the second probability 
factor.

In fact, the condition used above to obtain Eq.~(\ref{eq:P-classical}), 
$\inner{r_x(t_{\rm exp})}{r_y(t_{\rm exp})} \ll 1$ for $x \neq y$, is 
exactly what we expect.  Since the detector states $\bigl|\apparatusx\bigr>$ 
for different $x$ have different macroscopic configurations {\it and} 
since the Hamiltonian of the system is local, i.e.\ has the form of 
Eq.~(\ref{eq:QFT-local}), the future states corresponding to different 
$\bigl|\apparatusx\bigr>$ are almost orthogonal:%
\footnote{Note that this would not generally be the case if the relevant 
 object were microscopic, i.e.\ described by a Hilbert space with small 
 dimensions.  In that case, two different states could significantly 
 overlap (recohere) in the future, as the states $\ket{1}$ and $\ket{2}$ 
 in Eq.~(\ref{eq:12-evolve}).  Indeed, quantum interference effects 
 arise precisely because of such recoherence, which may occur easily 
 in a system with small Hilbert space dimensions.}
\begin{equation}
  \inner{\psi_x(t)}{\psi_y(t)} \sim \delta(x-y).
\label{eq:psi-ortho}
\end{equation}
(For example, photons scattered from the detector will have different 
configurations for different $x$~\cite{Joos:1984uk}.)  This implies 
that $\inner{r_x(t)}{r_y(t)} \sim \delta(x-y)$ because $\ket{\phi_x(t)}$ 
is only a very small subset of the entire degrees of freedom in 
$\ket{\psi_x(t)}$, leading to Eq.~(\ref{eq:P-classical}).

The argument given above, however, is not by itself sufficient to 
explain selection of the measurement basis, because the same analysis 
applies to any (not necessarily classical state) basis $z = \int\!dx\, 
r_{zx}\, x$ satisfying $\int\!dx\, r_{zx} r_{z'x} = \delta(z-z')$.  The 
selection happens because ``amplification'' occurs (only) in a particular 
basis (called quantum Darwinism)~\cite{q-Darwinism}.  Schematically,
\begin{equation}
  \ket{\psi_x(0)} \rightarrow \ket{\psi_x(t)} 
  \equiv \ket{x} \ket{X} \ket{\cal X} \cdots,
\label{eq:x-amplif}
\end{equation}
where the last expression implies that the same {\it classical} information, 
i.e.\ ``the double-slit experiment measured the electron at $x$,'' is 
available (independently) to many subcomponents of the system.  For 
example, for a superposition of two outcomes after the double-slit 
experiment, this gives
\begin{equation}
  \ket{\Psi(0)} = c_x \ket{\psi_x(0)} + c_y \ket{\psi_y(0)}
  \rightarrow c_x \left( \ket{x} \ket{X} \ket{\cal X} \cdots \right) 
    + c_y \left( \ket{y} \ket{Y} \ket{\cal Y} \cdots \right).
\label{eq:xy-amplif}
\end{equation}
Note that this is different from the cloning of quantum information, 
which would yield
\begin{equation}
  \ket{\Psi(0)} 
  \rightarrow \left( c_x \ket{x} + c_y \ket{y} \right) 
    \left( c_x \ket{X} + c_y \ket{Y} \right) 
    \left( c_x \ket{\cal X} + c_y \ket{\cal Y} \right) \cdots.
\label{eq:xy-copy}
\end{equation}
Namely, only selected information, that corresponding to classical 
states $\ket{\alpha_i}$, can be amplified.  The origin of the particular 
evolution in Eq.~(\ref{eq:xy-amplif}) is the special form of the 
evolution operator, Eq.~(\ref{eq:QFT-local})---the classical state 
basis is selected as a dynamical consequence of spacetime locality. 
While this is yet to be proven in the general case, analyses of simple 
quantum mechanical systems~\cite{q-Darwinism} strongly suggest it 
to be the case.

The properties of Eqs.~(\ref{eq:psi-ortho}) and (\ref{eq:xy-amplif}) 
for $\ket{\alpha_i}$ imply that if $n-1$ successive experiments are 
performed after the double-slit experiment, we obtain
\begin{eqnarray}
  P(A^{(n)}|{\rm exp}) &=& \sum_{A^{(n-1)}} \cdots \sum_{A^{(2)}} 
    \int\!\!dx\, P(x \,|\, \mbox{double-slit exp})\, 
\nonumber\\
  && {} \qquad \times P_x(A^{(2)} \,|\, \mbox{second exp}) \cdots\, 
    P_{x A^{(2)} \cdots A^{(n-1)}}(A^{(n)} \,|\, \mbox{$n$-th exp})
\label{eq:P-classical-2}
\end{eqnarray}
(up to negligible corrections, e.g., from $\inner{\psi_x(t)}{\psi_y(t)} 
\ll 1$ for $x \neq y$).  This is the classical probability formula.  The 
subscripts in $P$'s on the right-hand side indicate that the results of 
earlier experiments, e.g.\ $x$ of the double-slit experiment, are available 
independently to many successive experiments, just by accessing small 
subsets of the entire system.  This forms a crucial ingredient for 
classical objectivity~\cite{q-Darwinism}.  We emphasize that the 
basis of decomposition in Eq.~(\ref{eq:P-classical-2}) is determined 
{\it dynamically}.  In fact, the very {\it existence} of a special basis 
in which the classical formula of Eq.~(\ref{eq:P-classical-2}) is true 
is a dynamical consequence of spacetime locality.

The absence of the interference term in Eq.~(\ref{eq:P-classical-2}) 
implies that we can describe the system using ``decohered'' density matrix
\begin{equation}
  \rho = \frac{1}{\int\!dy\,|\psi_1(y) + \psi_2(y)|^2} 
    \int\!\!dx\, |\psi_1(x) + \psi_2(x)|^2 
    \Bigl\{ \ket{0} \bra{0} \Bigr\} \otimes \Big\{ \bigl|\apparatusx\bigr> 
    \bigl<\apparatusx\bigr| \Bigr\} \otimes \Bigl\{ 
    \bigl|\mbox{\large $R_{}$}\bigr> \bigl<\mbox{\large $R_{}$}\bigr| \Bigr\}
\label{eq:rho_fin}
\end{equation}
after the electron is measured in the double-slit experiment.  In particular, 
if we are interested only in physics after the electron is measured 
at a particular point $x$, then we need only keep the term containing 
$\bigl|\apparatusx\bigr>$, which is equivalent to using the ``collapsed'' 
state
\begin{equation}
  \ket{\Psi_{\rm col}} = \frac{\psi_1(x) + \psi_2(x)}{|\psi_1(x) + \psi_2(x)|} 
    \ket{0} \otimes \bigl|\apparatusx\bigr> 
    \otimes \bigl|\mbox{\large $R_{}$}\bigr>.
\label{eq:exp-fin-Cop}
\end{equation}

The argument presented here is the heart of the basis selection in 
describing any experimental result {\it in a way that our classical 
intuition is manifest}.  It also provides a real rationale behind 
Eq.~(\ref{eq:O_A-local}), which was chosen to be diagonal in the 
classical state basis $\ket{\alpha_i}$.  The information about classical 
configuration $\ket{\alpha_i}$ is that which can be amplified by 
the dynamical evolution---this is not very surprising given that 
$\ket{\alpha_i}$'s are deeply related to the locality basis states, 
which are determined by the form of the time evolution operator. 
Questions we ask are about information that can be objectively 
accessed by multiple physical processes (including what is stored in 
memory states of our brains), hence the form of Eq.~(\ref{eq:O_A-local}). 
Note that by integrating out $\bigl|\mbox{\large $R_{}$}\bigr>$ as well 
as all the histories after $t=0$, this reproduces the usual einselection 
criterion in the decoherence paradigm.  The argument here, however, 
makes it clear that the basis selection has nothing to do with the 
{\it ultimate} openness of the system---indeed, the present argument 
still applies even if the entire system is closed, being subject to 
thermalization and recurrences at later times.  The origin of the 
basis selection lies {\it entirely in the dynamics}---more specifically, 
the fact that the time evolution operator takes a special form of 
Eq.~(\ref{eq:QFT-local}) in the locality basis.

\subsection{Ordered observations require infinitely large Hilbert space}
\label{subsec:inf-needed}

Does the preceding argument ensure that the two features listed at the 
beginning of this section, (i) and (ii), are valid in any quantum system 
described by a local theory?  In other words, can we always consider that 
a sufficiently macroscopic measurement collapses the wavefunction to one 
of the possible states having a well-defined classical configuration?

The answer is yes, as long as we are interested only in dynamics at 
timescales much sorter than $t_{\rm th}$.  However, if the system is 
closed, and we are interested in arbitrarily long timescales, then the 
answer is nontrivial.  In this case the entire system thermalizes at 
$\sim t_{\rm th}$, after which it occasionally experiences rare fluctuations 
producing low entropy regions, and eventually comes back to a state 
that is arbitrarily close to the original state at timescale $t_* \sim 
e^{S_{\rm th}}$, where $S_{\rm th}$ is the thermal entropy of the system. 
This picture applies regardless of the details of the system, as long 
as quantum mechanical evolution is unitary and the initial condition 
is generic, which we assume throughout.  Since the process of producing 
low entropy fluctuations generically involves interferences between 
macroscopically different terms, replacing $\ket{\Psi_{\rm fin}}$ by 
the collapsed state $\ket{\Psi_{\rm col}}$ might seem to give an obviously 
wrong answer.  This conclusion, however, is too naive.

Consider a process in which a thermal state having a large coarse-grained 
entropy, $\sim S_{\rm th}$, fluctuates into a state with a low 
coarse-grained entropy and then evolves back to another thermal state:
\begin{equation}
  \bigl|\Psi_{\sim S_{\rm th}}\bigr>
\quad\rightarrow\quad
  \bigl|\Psi'_{S_{\rm low}}\bigr>
\quad\rightarrow\quad
  \bigl|\Psi''_{\sim S_{\rm th}}\bigr>,
\label{eq:fluctuate}
\end{equation}
where we have denoted the coarse-grained entropy of a state by the 
subscript ($S_{\rm low} \ll S_{\rm th}$).  One might think that the first 
part of this process, $\bigl|\Psi_{\sim S_{\rm th}}\bigr> \rightarrow 
\bigl|\Psi'_{S_{\rm low}}\bigr>$, looks like the ``{\it classical-to-quantum} 
transition,'' as it involves recoherence of macroscopically different 
configurations, thus contradicting observation.  This is, however, not 
true.  Because of the reversibility of quantum evolution, the process 
is given, with high probability, by the time reversal of a usual entropy 
increasing process~\cite{Aguirre:2011ac}:
\begin{equation}
  \bigl|\Psi_{\sim S_{\rm th}}\bigr> \rightarrow 
    \bigl|\Psi'_{S_{\rm low}}\bigr>
\quad\stackrel{\mbox{\footnotesize time reversal}}{\Longleftrightarrow}\quad
  \bigl|\bar{\Psi}'_{S_{\rm low}}\bigr> 
    \rightarrow \bigl|\bar{\Psi}_{\sim S_{\rm th}}\bigr>,
\label{eq:time-rev}
\end{equation}
where a state with a bar is the $CPT$ conjugate of the corresponding 
state without.  Since any physical observer is necessarily a part of 
the system, the two processes in Eq.~(\ref{eq:time-rev}) look identical 
to him/her---these are really time reversals of each other {\it including 
memory states of the observer}.  In particular, they both look like 
a standard thermalization process, involving the {\it quantum-to-classical} 
transition.%
\footnote{This statement seems evident to the author, given that the 
 dynamics is such that macroscopic correlations between subsystems increase 
 as entropy increases, and that ``perception'' is nothing but (classical) 
 memory states of physical computational devices, including our brains 
 (see e.g.~\cite{Schulman:2005}).  If it does not hold, however, it 
 would only strengthen our conclusion that the entire universe cannot 
 be a finite, closed system.}

\begin{figure}[t]
  \center{\includegraphics[scale=1.0]{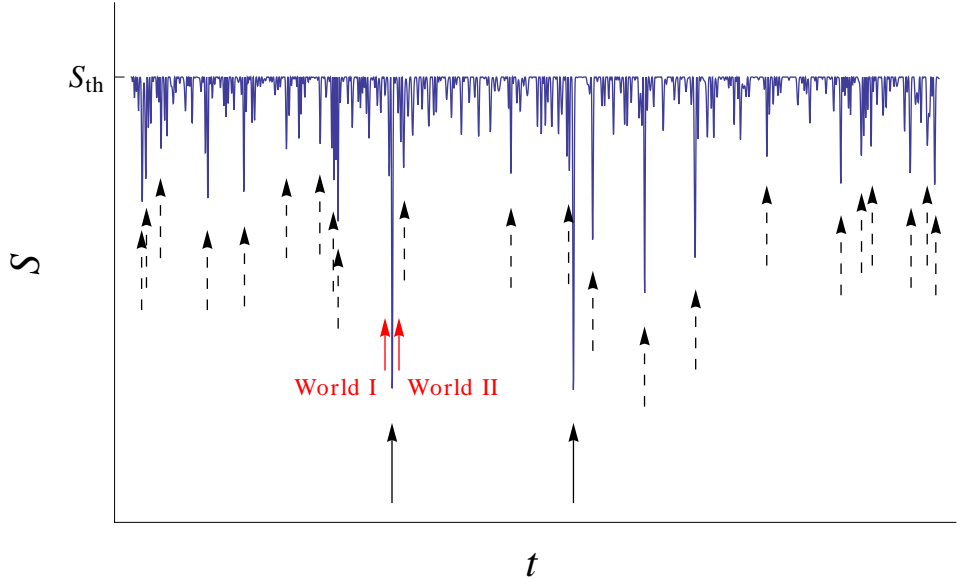}}
\caption{Entropy fluctuations in a closed quantum system can provide 
 ``observations,'' indicated by (both solid and dashed) arrows, with 
 each fluctuation corresponding to two ``regular worlds'' obeying the 
 second law of thermodynamics (as well as the standard rules of Copenhagen 
 quantum mechanics).  However, the vast majority of these fluctuations 
 correspond to random, irregular observations (dashed arrows), as 
 opposed to ordinary, ordered observations, which compose only a tiny 
 fraction of the fluctuations (solid arrows).}
\label{fig:fluct}
\end{figure}

In fact, if we follow the system for an arbitrarily long time, replacing 
$\ket{\Psi_{\rm fin}}$ by $\ket{\Psi_{\rm col}}$ at one time will lead 
only to a negligible error on final probabilities, obtained by a formula 
based on matrix elements (e.g.\ Eq.~(\ref{eq:probability-AB}), which 
will be discussed in detail in Section~\ref{sec:measure}).  In this 
case, the probability for a fluctuation such as Eq.~(\ref{eq:fluctuate}) 
to occur is simply proportional to the Boltzmann factor, multiplied by 
the number of microstates:\ $e^{-M_{\rm fluc}/T + S_{\rm low}}$ where 
$M_{\rm fluc}$ is the mass associated with the fluctuation and $T$ 
the temperature of the system.  For an ``observer,'' who can arise 
only as a part of such a fluctuation, the world always appears to obey 
properties (i) and (ii):\ the standard rules of Copenhagen quantum 
mechanics (as well as the second law of thermodynamics), as illustrated 
by two (red) arrows in Fig.~\ref{fig:fluct} labeled by ``World I'' 
and ``World II,'' respectively.

What is wrong, then, with the picture that the universe is a closed 
quantum mechanical system, with the history repeating with the period 
of order $t_*$?  The problem is that a fluctuation, which for internal 
observers appears as two ``regular worlds'' related by $CPT$, generically 
has ``initial conditions'' (i.e.\ $\bigl|\Psi'_{S_{\rm low}}\bigr>$ 
and $\bigl|\bar{\Psi}'_{S_{\rm low}}\bigr>$ in Eq.~(\ref{eq:time-rev})) 
that are {\it not} expected to be obtained by evolving the system from 
a state having a smaller coarse-grained entropy.  Indeed, according 
to standard equilibrium thermodynamics, the probability distribution 
of the ``initial conditions'' should follow $\propto e^{-F/T}$, where 
$F = M_{\rm fluc} - T S_{\rm low}$ is the free energy associated with 
the ``initial configurations'' (i.e.\ the configurations at the bottom 
of entropy dips).  This implies that if we consider processes that 
involve any ``perceptions,'' they will be overwhelmingly dominated 
by those observing random, irregular worlds, as opposed to a regular 
world obeying consistent laws of physics (because such random 
perceptions/consciousnesses arise much more easily from fluctuations; 
see Fig.~\ref{fig:fluct} for a schematic depiction), contradicting 
what we actually observe (assuming, of course, that we are ``typical'' 
perceptions).  This is nothing but the well-known Boltzmann brain 
problem~\cite{Page:2006dt} as applied to a general closed system.

Hence a description of the universe consistent with our observation, 
i.e.\ item (iii) listed at the beginning of this section, is obtained 
only if the thermalization timescale of the system, $t_{\rm th}$, is 
(much) larger than the timescale of interest $t$.  Since $t_* \sim 
e^{S_{\rm th}} \gg t_{\rm th}$ and $S_{\rm th} < \ln[{\rm dim}\,{\cal H}]$, 
where ${\rm dim}\,{\cal H}$ is the Hilbert space dimension of the entire 
system, this implies
\begin{equation}
  t < t_{\rm th} \ll t_* \sim e^{S_{\rm th}} < {\rm dim}\,{\cal H}.
\label{eq:large-Hilbert}
\end{equation}
(Note that the unit of time does not matter in the last inequality because 
$e^{S_{\rm th}}$ is double-exponentially large, with $S_{\rm th}$ being 
macroscopic.)  In particular, this implies that if we want to describe 
the entire history of the universe ($t \rightarrow \infty$), which we 
must do if quantum mechanical evolution is fundamentally unitary, then 
we need to take
\begin{equation}
  {\rm dim}\,{\cal H} = \infty,
\label{eq:Hilbert-inf}
\end{equation}
i.e.\ {\it the Hilbert space describing the quantum universe must 
be infinitely large}.

As we will see in Section~\ref{sec:Hilbert-QG}, the condition of 
Eq.~(\ref{eq:Hilbert-inf}) is satisfied in the eternally inflating 
multiverse because the multiverse evolves asymptotically to a supersymmetric 
Minkowski (or singularity) world, which contains an infinite number of 
states.  This completes an ultimate picture for quantum measurement---{\it 
a quantum measurement is a process in which a coherence existing in a 
(microscopic) system is dissipated into larger systems, ultimately into 
states in a supersymmetric Minkowski (or singularity) world}.  During 
this process, classical information about the measurement result (e.g.\ 
$x$ of the double-slit experiment) is amplified in {\it each component} 
of the multiverse state.  Because of an infinitely large coarse-grained 
entropy of Minkowski (and singularity) space, recoherence of macroscopically 
different worlds does not occur.  {\it The ordered, classical world we 
see is a consequence of the fact that the universe (multiverse) is in 
an infinite-dimensional representation of the time evolution operator.} 
The evolution itself, however, still obeys the laws of quantum mechanics; 
in particular, it is deterministic and unitary.

\section{Spacetime Locality in Theories with Gravity}
\label{sec:locality-grav}

We have seen that spacetime locality plays a crucial role in quantum 
measurement processes.  In theories with gravity, however, this 
property---i.e.\ that the time evolution operator takes a special 
form of Eq.~(\ref{eq:QFT-local})---is not automatically guaranteed. 
In particular, if we take wrong hypersurfaces to quantize the system, 
theories are not local {\it even at distances much larger than the quantum 
gravity scale}.  Historically, this issue appeared first in the study 
of black holes, but it is much more general and provides an important 
constraint on how to define quantum states in theories with gravity.

In the rest of the paper, we derive the structure of the Hilbert space 
describing the entire quantum universe, starting with the well-known 
discussion on quantum mechanics of black holes.  A key ingredient is {\it 
to fix all the redundancies} and {\it to eliminate all the overcountings} 
of a general relativistic description of nature.  In this broader 
context, black hole complementarity arises as a part of the general 
transformation of the Hilbert space associated with a change of the 
``reference frame.''  This transformation can be regarded as an extension 
of the Lorentz/Poincar\'{e} transformation in the quantum gravitational 
context.  With this Hilbert space, the probabilities are defined following 
the argument presented in previous sections, with an important extension 
to make physics invariant under time reparameterization (as required 
in theories with gravity).

The basic results presented in the following, such as the structure 
of the Hilbert space and the probability formula, were obtained mostly 
in Ref.~\cite{Nomura:2011dt}.  There are, however, some important 
refinements here, including the treatment of spacetime singularities 
and a useful probability formula that applies in many practical cases. 
We also provide a clearer argument leading to the results, and discuss 
their meaning especially in the context of physical measurement processes. 
As emphasized in Ref.~\cite{Nomura:2011dt}, these measurements can 
be either regarding global properties of the universe or outcomes of 
particular experiments.  This, therefore, provides complete unification 
of the eternally inflating multiverse and the many worlds interpretation 
of quantum mechanics.

\subsection{Black hole complementarity}
\label{subsec:BH-compl}

Here we review black hole 
complementarity~\cite{Susskind:1993if,Susskind:2005js}, which we assume 
gives the correct description of black hole physics.  We focus particularly 
on aspects relevant to our later discussions.

Consider a traveler falling into a black hole, carrying some information. 
From the point of view of a distant observer, this traveler is absorbed 
into the horizon, which has an extremely high local temperature.  (The 
temperature of radiation arriving at the distant observer is enormously 
redshifted, and is given by the standard formula $1/8\pi G_N M_{\rm BH}$.) 
Assuming that quantum mechanics is valid, the original information carried 
by the traveler must be stored at the horizon, which will eventually come 
out as Hawking radiation.  In the limit that the black hole is very large, 
i.e.\ a static black hole, this implies that the evolution of the system 
is unitary on the Hilbert space
\begin{equation}
  {\cal H}_{\rm BH}^{\rm (distant)} = 
  {\cal H}_{\rm horizon} \otimes {\cal H}_{\rm outside},
\label{eq:H-distant}
\end{equation}
where ${\cal H}_{\rm horizon}$ and ${\cal H}_{\rm outside}$ represent the 
Hilbert space factors associated with the degrees of freedom {\it on} and 
{\it outside} the horizon, respectively.  (Strictly speaking, the horizon 
here means the stretched horizon, which is $\sim l_s$ away from the 
mathematical horizon, where $l_s$ is the string length.)

On the other hand, from the point of view of the falling traveler, there 
is nothing special about the horizon, and physics is described in the 
Hilbert space
\begin{equation}
  {\cal H}_{\rm BH}^{\rm (falling)} = 
  {\cal H}_{\rm inside} \otimes {\cal H}_{\rm outside},
\label{eq:H-falling}
\end{equation}
where ${\cal H}_{\rm inside}$ is the Hilbert space associated with the 
degrees of freedom {\it inside} the horizon.  Again, assuming the validity 
of quantum mechanics, the evolution of the system is unitary on the 
above Hilbert space (until the singularity is hit).

Now, let us consider the fate of the information originally carried 
by the traveler.  From the distant observer's viewpoint, elements of 
Eq.~(\ref{eq:H-distant}) will be mapped, after the back hole evaporates, 
into those of the Hilbert space associated with spacetime without the 
black hole:
\begin{equation}
  {\cal H}_{\rm horizon} \otimes {\cal H}_{\rm outside} 
    \mapsto {\cal H}_{\rm after\,\,evaporation}.
\label{eq:H-distant-map}
\end{equation}
(A more complete treatment of the Hilbert space for dynamically evolving 
spacetime will be given in Section~\ref{sec:Hilbert-frame}.)  The 
information is then first in ${\cal H}_{\rm horizon}$, and later in 
${\cal H}_{\rm after\,\,evaporation}$ as subtle quantum correlations 
in Hawking radiation.  From the falling traveler's viewpoint, on the 
other hand, this information is in ${\cal H}_{\rm inside}$.  A problem 
arises when we mix these two viewpoints in the global spacetime picture. 
In this picture, we can draw spacelike hypersurfaces---often called 
nice slices---on which the information exists {\it both} in Hawking 
radiation {\it and} inside spacetime.  From a general relativistic point 
of view, there is nothing wrong with defining states on such hypersurfaces. 
This, however, leads to contradiction with the laws of quantum mechanics, 
specifically the no-cloning theorem~\cite{Wootters:1982zz}, which says 
that quantum information cannot be faithfully duplicated.

Black hole complementarity asserts that the problem arises because we 
have taken the global viewpoint that does not have any operational meaning. 
Indeed, because of the existence of the horizon, no physical observer can 
obtain the {\it same} information from inside region {\it and} Hawking 
radiation~\cite{Susskind:1993mu}.  This implies that if quantum mechanics 
is defined on equal-time hypersurfaces that pass through both the inside 
and outside information, e.g.\ on nice slices, then the low energy {\it 
theory} (not just states) must be non-local in such a way that these 
spatially separated degrees of freedom are not independent.  (For 
suggestive calculations in string theory consistent with this picture, 
see Refs.~\cite{Lowe:1995ac,Lunin:2001jy}.)  Alternatively, if we want 
to keep locality in our low energy description of nature (which we 
do), then the Hilbert space should be restricted to the one associated 
with appropriate spacetime regions, e.g.\ Eq.~(\ref{eq:H-distant}) 
or (\ref{eq:H-falling}).  Namely, {\it including both the inside 
spacetime region and Hawking radiation in a single description is 
overcounting}---they cannot coexist as independent degrees of freedom 
in a single component of a quantum state.

\subsection{Quantum states are defined in restricted spacetime regions}
\label{subsec:single-obs}

How should we then define quantum states without sacrificing locality 
of the theory at distances larger than the quantum gravity scale? 
In Ref.~\cite{Nomura:2011dt}, it is proposed that:
\begin{itemize}
\item
The states are defined on the past light cone bounded by the (stretched) 
apparent horizon.  Here, the apparent horizon is defined as a surface on 
which at least one pair of orthogonal null congruences have zero expansion; 
in particular, the local expansion of past directed light rays emitted 
from the tip to form the past light cone turns from positive to negative 
there.  The degrees of freedom exist {\it both} inside {\it and} on the 
horizon, and the system is described as viewed from a local Lorentz frame 
at the tip of the light cone.%
\footnote{Defining states on the past light cone is not absolutely 
 necessary, although it provides the simplest formulation of the framework. 
 An alternative possibility is to use spacelike hypersurfaces foliating 
 the causal patch and define states on them in the region inside and 
 on the (stretched) apparent horizon.  Because the numbers of degrees of 
 freedom on the past light cone and on a spacelike hypersurface bounded 
 by the same horizon are the same in our setup (due to the spacelike 
 projection theorem~\cite{Bousso:1999xy}), the Hilbert space dimensions 
 in both cases are identical. \label{ft:spacelike}}
\item
The definition above provides the simplest way of avoiding the overcounting 
of the type described in Section~\ref{subsec:BH-compl}, making the time 
evolution operator local at distances larger than the quantum gravity 
scale.  The evolution of a quantum state is deterministic and unitary 
in this Hilbert space (until spacetime singularities are hit; see 
Section~\ref{subsec:singularities}).
\end{itemize}
The use of the apparent horizon in the above definition implies that we 
are selecting a ``reference frame'' from which we describe physics---indeed, 
the light cone by definition must be associated with some point in spacetime. 
What does that mean?  We will address this question in the next section, 
where we will also discuss the evolution of a quantum state in general 
dynamical spacetime.  Here we simply focus on the issue of unitarity 
in fixed background geometries.

Let us consider de~Sitter space.  Following the above definition, the 
states on this background are given on the past light cone of some fixed 
center $p$.  The structure of the Hilbert space, of which these states 
are elements, is thus
\begin{equation}
  {\cal H}_{\rm dS} = {\cal H}_{\rm dS,\, horizon} 
    \otimes {\cal H}_{\rm dS,\, inside},
\label{eq:H-dS}
\end{equation}
where ${\cal H}_{\rm dS,\, horizon}$ and ${\cal H}_{\rm dS,\, inside}$ 
are the Hilbert space factors associated with the degrees of freedom on 
and inside the stretched horizon, respectively.  On a fixed de~Sitter 
background, the evolution of a state is unitary in the space of 
Eq.~(\ref{eq:H-dS}).  In particular, information leaving the horizon 
of $p$ in the global spacetime picture is regarded as being stored 
in ${\cal H}_{\rm dS,\, horizon}$, which can later be sent back to 
${\cal H}_{\rm dS,\, inside}$ in the form of subtle quantum correlations 
in Hawking radiation.  Note that, in this description, a physical observer 
need not be at $p$, which simply plays the role of ``the origin of the 
coordinates.''  Indeed, since the information in Hawking radiation spreads 
non-locally over the space, gathering it requires physical processes 
at locations other than the origin (and therefore, these processes feel 
local temperatures higher than the Gibbons-Hawking temperature $H/2\pi$, 
where $H$ is the Hubble parameter).

\begin{figure}[t]
\begin{center}
\begin{picture}(280,205)(0,-150)
  \DashLine(0,0)(100,0){3}
  \Line(100,0)(140,40) \Line(99.3,0)(140,40.7)
  \Line(140,40)(180,0) \Line(140,40.7)(180.7,0)
  \DashLine(180,0)(280,0){3}
  \Line(100,0)(134,-34) \Line(99.3,0)(134,-34.7)
  \CArc(128.34,-39.66)(8,0,45) \CArc(127.84,-39.66)(8,0,45)
  \Line(180,0)(146,-34) \Line(180.7,0)(146,-34.7)
  \CArc(151.66,-39.66)(8,135,180) \CArc(152.16,-39.66)(8,135,180)
  \DashLine(140,0)(120,-20){2} \Line(120,-20)(10,-130)
  \DashLine(140,0)(160,-20){2} \Line(160,-20)(270,-130)
  \DashCArc(-121.63,-261.63)(370,40.5,45){2}
  \CArc(-121.63,-261.63)(370,20.84,40.5)
  \LongArrow(140,-130)(140,39) \Text(140,-135)[t]{$p$}
  \LongArrow(210,-130)(230,-1) \Text(233,-8)[lt]{\scriptsize traveler}
  \Photon(155,-45)(200,-90){1}{6} \Line(154,-44)(155,-45)
  \Line(154,-44)(163.90,-48.24) \Line(154,-44)(158.24,-53.90)
  \Text(182,-80)[r]{\footnotesize Hawking}
  \Text(189,-89)[r]{\footnotesize radiation}
  \Line(179,-21)(214,-56) \Line(181,-19)(216,-54)
  \Line(177.5,-17.5)(187.40,-21.74) \Line(177.5,-17.5)(181.74,-27.40)
  \Text(212,-14)[r]{\footnotesize direct}
  \Text(219,-23)[r]{\footnotesize signal}
\end{picture}
\caption{In the situation where a Minkowski bubble forms in a meta-stable 
 de~Sitter vacuum, there are two possible ways to retrieve information 
 carried away from the horizon of $p$ by some traveler:\ from Hawking 
 radiation and from a direct signal.  The conditions for successful 
 information retrieval from these sources, however, are mutually 
 incompatible, prohibiting faithful duplication of quantum information.}
\label{fig:gedanken}
\end{center}
\end{figure}
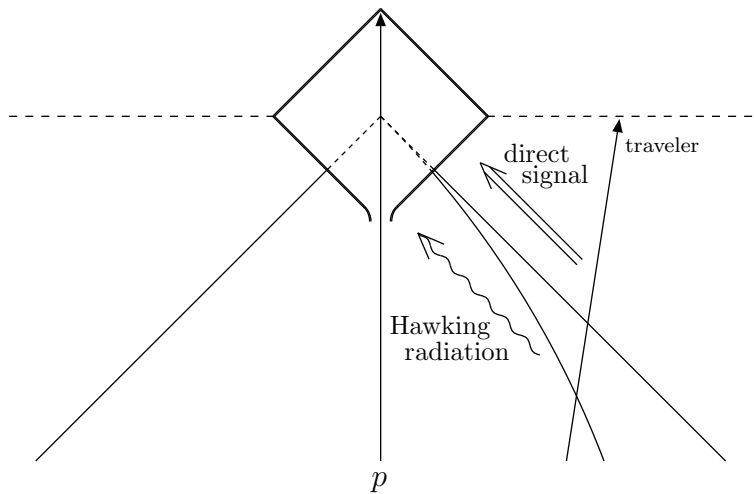

A nontrivial consistency check of this picture was given in Appendix~C 
of Ref.~\cite{Nomura:2011dt}, which we briefly reproduce here.  Let us 
consider a semi-classical geometry in which a Minkowski bubble forms in 
the future of a meta-stable de~Sitter vacuum.  We assume that the reference 
point $p$ enters into the Minkowski bubble at a time $t_{\rm nuc}$ in 
the flat coordinates of the de~Sitter space.  Since the information 
retrieval time from the de~Sitter horizon is $t_{\rm ret} \simeq H^{-1} 
\ln(H^{-1}/l_P)$, where $l_P$ is the Planck length, we need to have
\begin{equation}
  t_{\rm nuc} \simgt t_{\rm ret} 
  \simeq \frac{1}{H} \ln\left(\frac{1}{l_P H}\right),
\label{eq:nuc-ret}
\end{equation}
in order to retrieve any information {\it from Hawking radiation} that 
was ``carried away'' by a traveler who left the horizon at time $t = 0$. 
On the other hand, we may consider a process in which an observer (who 
sits within $p$'s horizon) obtains information {\it directly} from the 
traveler after he/she enters the Minkowski bubble, by receiving some 
signal; see Fig.~\ref{fig:gedanken}.  It turns out that this succeeds 
only if the bubble nucleation occurs early enough:
\begin{equation}
  t_{\rm nuc} \simlt \frac{1}{H} \ln\left(\frac{1}{l_P H}\right),
\label{eq:nuc-direct}
\end{equation}
given that the traveler needs a Planck time to send a bit of information. 
We find that Eq.~(\ref{eq:nuc-ret}) and Eq.~(\ref{eq:nuc-direct}) are 
(barely) {\it in}compatible---no physical observer can receive duplicate 
quantum information from Hawking radiation {\it and} from the direct signal. 
Note that a physical observer can obtain information from Hawking radiation 
only before entering the Minkowski bubble, since the information is spread 
non-locally throughout the de~Sitter space.  Although some of the 
Hawking radiation quanta originated in the parent de~Sitter vacuum 
could reach the Minkowski bubble across the bubble wall, this would 
not be enough to transmit the information to the observer.

The examples considered above clearly demonstrate that the restriction 
of spacetime regions is crucial to keep locality of the low energy theory 
while being consistent with quantum mechanics.  Assuming that information 
absorbed into the de~Sitter horizon can be retrieved later (which is 
necessary for stable de~Sitter space to be regarded as a closed system, 
as suggested by the holographic principle), quantum states cannot be 
defined on hypersurfaces that pass through both Hawking radiation and 
outside spacetime containing the same information.  (Such a description 
would require a low energy theory to be non-local at distances larger 
than the quantum gravity scale.)  This situation is analogous to that 
in the black hole case.  The only difference is that the de~Sitter 
horizon is ``observer dependent'':\ its location changes depending 
on from whose point of view the system is described.

In Ref.~\cite{Nomura:2011dt}, the definition of quantum states considered 
above was stated as:\ the system is described from the viewpoint of 
a single ``observer'' (geodesic).  Here we phrase the same thing as:\ 
{\it physics should be described using a single reference frame.} 
This might capture the essential physics better.  We have seen that 
choosing a reference frame involves a restriction of spacetime regions 
in which quantum states are defined.  In the next section, we see the 
procedure of defining states in more detail from this perspective.

\section{Importance of Fixing a Reference Frame}
\label{sec:Hilbert-frame}

In this section we determine the structure of the Hilbert space for 
dynamical spacetime, Eq.~(\ref{eq:multiverse-H}).  (The Hilbert space 
for full quantum gravity, including spacetime singularities, will be 
discussed in the next section.)  We emphasize the importance of fixing 
a reference frame.  We also consider the effect of changing the reference 
frame, and see that it leads to a transformation in the Hilbert space that 
can be viewed as an extension of the Lorentz/Poincar\'{e} transformation 
in the context of quantum gravity.

\subsection{Fixing a gauge---physics should be described in a single 
 reference frame}
\label{subsec:fix-gauge}

What are observables in physical theories?  They should be ``gauge 
invariant,'' i.e.\ quantities that do not depend on arbitrary 
parameterizations of the system corresponding to the redundancy of 
the description.  In theories with gravity, the coordinatization of 
spacetime is precisely one such parameterization, so it might be thought 
that only observables are certain global quantities, e.g.\ the ones 
associated with the topology of spacetime.  This is not true---causal 
relations among events are invariant under general coordinate 
transformations, and thus are physically observable.

It is well known that to do Hamiltonian quantum mechanics, all the gauge 
redundancies must be fixed.  A theory of gravity has huge redundancies 
associated with general coordinate transformations:\ general covariance. 
Indeed, without gauge fixing, the Hamiltonian one would obtain is zero, 
reflecting invariance under local time translations.  The definition of 
the states described in the previous section provides a simple way to 
fix these redundancies and extract causal relations among the events. 
The origin of a special point $p$ (i.e.\ the tip of the past light cone 
used to define states) is now clear---it arises from the fact that the 
theory is invariant under local spatial translations and that we need 
to fix the resulting redundancies.  By choosing a local Lorentz frame 
with the origin at $p$, all the redundancies associated with $p$ are 
fixed.  While this prescription by itself does not completely determine 
the gauge for general covariance, fixing the residual ones, coming from 
coordinate transformations on the past light cone, is simple conceptually 
and gives only minor effects on the overall picture.

Together with the restriction of spacetime within the (stretched) apparent 
horizons discussed in Section~\ref{subsec:single-obs}, this comprises 
the statement in Ref.~\cite{Nomura:2011dt} that ``physics is described 
from the viewpoint of a single observer.''  The choice of the local 
Lorentz frame at $p$ implies that the tip of the light cone follows 
a geodesic at the semi-classical level.  The overcounting of the type 
encountered before does not arise, and the time evolution operator 
is local at large distances.%
\footnote{Strictly speaking, there are apparent non-localities associated 
 with gauge fixing and with the fact that we quantize the system on null 
 hypersurfaces.  These non-localities, however, are different from the 
 ``real'' non-locality discussed in Section~\ref{sec:locality-grav}. 
 For example, the former disappears in physical quantities, and the 
 latter is associated with propagation of massless particles, a process 
 that is local in spacetime (and so it does not arise if we instead use 
 spacelike hypersurfaces; see footnote~\ref{ft:spacelike}).}
Here we summarize all these as:\ {\it physics should be described in 
a single reference frame if we want to keep locality of the theory 
at distances larger than the quantum gravity scale}.

\subsection{Hilbert space for dynamical spacetime---analogy with Fock space}
\label{subsec:Hilbert}

We now construct the Hilbert space for dynamical spacetime, following the 
discussion so far.  To do so, it is instructive to draw a close analogy 
with the construction of the Hilbert space in usual (non-gravitational, 
non-conformal) quantum field theory.

Consider a non-gravitational quantum field theory in which asymptotic states 
contain a single species of particles described by creation/annihilation 
operators $a^\dagger_{{\bf p},s}$/$a_{{\bf p},s}$, where ${\bf p}$ and 
$s$ are the momentum and spin indices, respectively.  (An extension to 
the case of multiple species is straightforward.)  The Hilbert space of 
the theory is then (isomorphic to) the Fock space
\begin{equation}
  {\cal H} = \bigoplus_{n = 0}^{\infty} {\cal H}_{\rm 1P}^{\otimes n},
\label{eq:QFT-H}
\end{equation}
where ${\cal H}_{\rm 1P}^{\otimes n}$ represents the $n$-particle Hilbert 
space given by
\begin{equation}
  {\cal H}_{\rm 1P}^{\otimes 0} \equiv \ket{0},
\qquad
  {\cal H}_{\rm 1P}^{\otimes 1} 
    = \left\{\, a^\dagger_{{\bf p},s} \ket{0} \,\right\},
\qquad
  {\cal H}_{\rm 1P}^{\otimes 2} 
    = \left\{\, a^\dagger_{{\bf p_1},s_1} a^\dagger_{{\bf p_2},s_2} 
      \ket{0} \,\right\},
\qquad
  \cdots.
\label{eq:QFT-H_1P}
\end{equation}
With generic interactions, a state in ${\cal H}$ in Eq.~(\ref{eq:QFT-H}) 
evolves across different components ${\cal H}_{\rm 1P}^{\otimes n}$, i.e., 
the time evolution operator allows for a process changing the particle 
number.  For example, in the Standard Model, collision of an electron and 
a positron having well-defined momenta/spins at $t = -\infty$ leads to
\begin{equation}
  \Psi(t = -\infty) = \ket{e^+ e^-}
\quad\rightarrow\quad
  \Psi(t = +\infty) = c_e \ket{e^+ e^-} + c_\mu \ket{\mu^+ \mu^-} + \cdots,
\label{eq:QFT-evolution}
\end{equation}
where we have expanded the state $\Psi(t)$ in terms of the Fock-space 
states, which is appropriate for $t \rightarrow \pm \infty$ when 
interactions are weak.  (We have suppressed momentum and spin indices.) 
Note that Eq.~(\ref{eq:QFT-evolution}) should {\it not} be interpreted 
that the initial $e^+e^-$ state evolves probabilistically into different 
states $\ket{e^+ e^-}$, $\ket{\mu^+ \mu^-}$, and so on.  According 
to the laws of quantum mechanics, the evolution of the state $\Psi(t)$ 
is {\it deterministic}---it simply evolves into a unique state 
$\Psi(t = +\infty)$ which contains components $\ket{e^+ e^-}, 
\ket{\mu^+ \mu^-}, \cdots$ when expanded in Fock-space states.

The situation in quantum gravity is analogous.  We first need to fix 
the Hilbert space basis to discuss states unambiguously.  We assume 
that, with a fixed local Lorentz frame associated with a fixed reference 
point $p$, we have a set of local operators at low energies; specifically, 
we have a set of quantum fields $\phi_i({\bf x})$ defined on the past 
light cone of $p$.  (These fields do not depend on time as we take 
the Schr\"{o}dinger picture.)  This can provide ``meaning'' to the 
states according to the responses to these field operators, and we 
can now construct states using the language of, e.g., spacetime points, 
which are already fixed by the operator algebra (see the discussion in 
Section~\ref{subsec:locality-basis}).

Let us consider Hilbert space ${\cal H}_{\cal M}$ corresponding to 
a set of fixed semi-classical geometries ${\cal M} = \{ {\cal M}_i \}$, 
which are defined on the past light cone of $p$ and have the same 
(stretched) apparent horizon $\partial {\cal M}$ (in the sense that 
they lead to the same internal geometry of $\partial M$).  Note 
that if $\partial {\cal M}$ is $d-2$ dimensional, the corresponding 
${\cal M}_i$'s (which are then $d-1$ dimensional) represent $d$ 
dimensional spacetime.  As discussed before, the states are defined 
on the past light cone bounded by the horizon; specifically, the states 
on these geometries form Hilbert space
\begin{equation}
  {\cal H}_{\cal M} = {\cal H}_{{\cal M}, {\rm bulk}} 
    \otimes {\cal H}_{{\cal M}, {\rm horizon}},
\label{eq:ST-H_M}
\end{equation}
where ${\cal H}_{{\cal M}, {\rm bulk}}$ and ${\cal H}_{{\cal M}, 
{\rm horizon}}$ represent Hilbert space factors associated with the 
degrees of freedom inside and on $\partial {\cal M}$, respectively. 
The product space structure is dictated by locality.  What do we 
know about ${\cal H}_{{\cal M}, {\rm bulk}}$ and ${\cal H}_{{\cal M}, 
{\rm horizon}}$?  The covariant entropy bound~\cite{Bousso:1999xy} 
suggests that the dimension of ${\cal H}_{{\cal M}, {\rm bulk}}$ is 
$\exp({\cal A}_{\partial {\cal M}}/4)$, where ${\cal A}_{\partial 
{\cal M}}$ is the area of the horizon $\partial {\cal M}$  in Planck 
units, and the standard horizon entropy implies that the dimension 
of ${\cal H}_{{\cal M}, {\rm horizon}}$ is the same, so
\begin{equation}
  {\rm dim}\,{\cal H}_{\cal M} 
  = {\rm dim}\,{\cal H}_{{\cal M}, {\rm bulk}} \times
    {\rm dim}\,{\cal H}_{{\cal M}, {\rm horizon}} 
  = \exp\left(\frac{{\cal A}_{\partial {\cal M}}}{2}\right).
\label{eq:dim-H_M}
\end{equation}
The fact that the maximum number of degrees of freedom (i.e.\ the logarithm 
of the dimension of the Hilbert space) scales with the area, rather 
than the volume, is a manifestation of the holographic principle.

Analogously to the case of non-gravitational quantum field theory, 
Eq.~(\ref{eq:QFT-H}), the full Hilbert space of dynamical spacetime 
is (isomorphic to) the direct sum of the Hilbert spaces for different 
${\cal M}$'s:
\begin{equation}
  {\cal H} = \bigoplus_{\cal M} {\cal H}_{\cal M},
\label{eq:ST-H}
\end{equation}
so that
\begin{equation}
  {\rm dim}\,{\cal H} 
    = \sum_{\cal M} {\rm dim}\,{\cal H}_{\cal M} 
    = \sum_{\cal M} 
      \exp\left(\frac{{\cal A}_{\partial {\cal M}}}{2}\right).
\label{eq:dim-ST-H}
\end{equation}
Since ${\rm dim}\,{\cal H}_{\cal M}$ are integers, ${\cal A}_{\partial 
{\cal M}}$ are quantized.  This suggests that $\sum_{\cal M}$ in the 
above equation may indeed be a discrete sum in the full quantum theory, 
although the argument itself does not prohibit a continuous degeneracy 
for a fixed ${\cal A}_{\partial {\cal M}}$.  Note that the dimension 
in Eq.~(\ref{eq:dim-ST-H}) includes that of ``matter'' degrees 
of freedom, i.e.\ excitations associated with the quantum fields 
$\phi_i({\bf x})$.

We emphasize that defining the states in ${\cal H}_{{\cal M}, {\rm bulk}}$ 
need not require a fixed background space {\it a priori}; rather, in a 
more complete framework, spacetime interpretation of the states would arise 
as a result of the algebra of quantum field operators, $\phi_i({\bf x})$, 
and the responses of the states to these operators.  There is a good 
feature in our framework in pursuing such a construction:\ since our 
states represent regions within the apparent horizon, on which the local 
expansion of light rays forming the past light cone of $p$ turns from 
positive to negative, the cross sectional area for a set of past directed 
light rays emanating from $p$ always increases as they move away from 
$p$.  This implies that these light rays do not form caustics, allowing 
for an unambiguous coordinatization.  More work, however, is needed to 
make this construction more precise.

The general evolution of a state in dynamical spacetime is unitary 
{\it in the full Hilbert space ${\cal H}$ in Eq.~(\ref{eq:ST-H})}, but 
not in each ${\cal H}_{\cal M}$.  The evolution, therefore, generically 
leads to the multiverse (or quantum many worlds) picture:
\begin{equation}
  \Psi(t = t_0) = \ket{\Sigma}
\quad\rightarrow\quad
  \Psi(t) = \sum_i c_i(t) \ket{\mbox{(cosmic) configuration $i$}},
\label{eq:multiverse-evolution}
\end{equation}
where $\ket{\Sigma}$ is an initial state at $t = t_0$, e.g.\ an eternally 
inflating (metastable de~Sitter) state in ${\cal H}_{{\cal M} = {\rm dS}}$, 
while the sum in the final state $\Psi(t)$ runs over states in different 
${\cal H}_{\cal M}$, giving a superposition of macroscopically different 
worlds (universes).  Quantum field theory on a fixed gravitational 
background corresponds to a very special case in which transitions between 
(some of the) ${\cal H}_{\cal M}$ are prohibited.  This is analogous 
to the nonrelativistic limit of usual quantum field theory, in which 
transitions between different ${\cal H}_{\rm 1P}^{\otimes n}$ do not 
occur.  The underlying dynamics for gravity, however, is much more 
general---the time evolution operator allows for ``hopping'' between 
different components ${\cal H}_{\cal M}$ in Eq.~(\ref{eq:ST-H}).  The 
semi-classical evolution in which the area of the apparent horizon 
changes is precisely a succession of such processes.

\subsection{``Reference frame dependence'' of the concept of spacetime}
\label{subsec:frame-change}

What happens if we change the reference frame, e.g.\ by a spatial 
translation or boost?  As in any symmetry transformation, this operation 
must be represented by a unitary transformation in Hilbert space. 
In particular, if we focus on histories before any component of the 
state hitting spacetime singularities (the effect of which will be 
discussed in Section~\ref{subsec:singularities}), then it must be 
represented entirely in the Hilbert space ${\cal H}$ in Eq.~(\ref{eq:ST-H}), 
{\it but not necessarily in each component ${\cal H}_{\cal M}$}.  Namely, 
the transformation can mix elements in different ${\cal H}_{\cal M}$. 
Moreover, even if the transformation maps all the elements in 
${\cal H}_{\cal M}$ onto themselves for some ${\cal M}$, there is 
no reason that it should not mix the degrees of freedom associated with 
${\cal H}_{{\cal M}, {\rm bulk}}$ and ${\cal H}_{{\cal M}, {\rm horizon}}$.

Let us consider a state $\ket{\Psi(t)}$ that represents the entire quantum 
universe, which we call {\it the multiverse state}.  Suppose that at 
some time $t$, the multiverse state is expanded in terms of the locality 
basis states $\ket{\kappa_m}$ that are elements of ${\cal H}$ in 
Eq.~(\ref{eq:ST-H}):
\begin{equation}
  \ket{\Psi(t)} = \sum_{m=1}^{{\rm dim}\,{\cal H}} c_m(t) \ket{\kappa_m},
\label{eq:Psi-multiverse}
\end{equation}
where $\inner{\kappa_m}{\kappa_n} = \delta_{mn}$.  The parameter $t$ 
here is introduced to describe the evolution of $\ket{\Psi(t)}$:\ 
$\ket{\Psi(t_1)} = e^{-iH(t_1-t_2)} \ket{\Psi(t_2)}$, where $H$ is the 
time evolution operator (i.e.\ gauge-fixed Hamiltonian) for the entire 
quantum universe.  Since the introduction of $t$ is arbitrary, the final 
physical predictions should not depend on this particular parameterization. 
(This is ensured by general covariance, which includes invariance under 
time reparameterization; see Section~\ref{subsec:Born} for more detail.) 
A useful choice for $t$ is the proper time at $p$, although any other 
monotonic parameterization works as well at the cost of (potentially) 
making the explicit form of $H$ complicated.

In $d$ spacetime dimensions, a change of the reference frame can be specified 
by $d(d+1)/2$ parameters $\{ r_i, \eta_i, \theta_{[ij]}, t\}$: $d-1$ 
spatial translations $r_i$, $d-1$ boosts $\eta_i$, and $(d-1)(d-2)/2$ 
rotations $\theta_{[ij]}$, performed at time $t$, where $i = 1,\cdots,d-1$. 
These correspond to the freedom in electing a local inertial frame in 
spacetime, from which we view the system.  Let us consider the Hilbert 
(sub)space corresponding to de~Sitter space:\ Eq.~(\ref{eq:H-dS}) where 
${\cal H}_{\rm dS,\, inside} \equiv {\cal H}_{{\cal M}={\rm dS,\, bulk}}$ 
and ${\cal H}_{\rm dS,\, horizon} \equiv {\cal H}_{{\cal M}={\rm dS,\, 
horizon}}$.  Consider a state in this space represented as
\begin{equation}
  \ket{\psi} = \ket{{\rm excited}}_{\rm inside} \otimes \ket{0}_{\rm horizon}
\label{eq:dS-before}
\end{equation}
at time $t$, where $\ket{{\rm excited}}_{\rm inside}$ implies that there 
is an object within the horizon, and $\ket{0}_{\rm horizon}$ is (one of) 
the ground state(s) for the horizon degrees of freedom.  If we change 
the reference frame by performing a large spatial translation (larger 
than the Hubble radius) at this time, then the state transforms as
\begin{equation}
  \ket{\psi} \quad
  \stackrel{\mbox{\footnotesize translation $T$ at $t$}}{\longrightarrow}
  \quad U_T \ket{\psi}
  = \ket{0}_{\rm inside} \otimes \ket{{\rm excited}}_{\rm horizon},
\label{eq:dS-after}
\end{equation}
where $\ket{0}_{\rm inside}$ represents (one of) the vacuum state(s) within 
the horizon, and $\ket{{\rm excited}}_{\rm horizon}$ an excited state 
for the horizon degrees of freedom.  This provides a simple example in 
which degrees of freedom in the bulk and on the horizon are mixed under 
a change of the reference frame.

A more drastic situation may occur when there is a black hole.  Consider 
a state $\ket{\phi}$ in ${\cal H}_{\cal M}$ at time $t$, where ${\cal M}$ 
is a spacetime containing a black hole but with the reference point 
$p$ staying outside the horizon, i.e. ${\cal H}_{\cal M} \subset 
{\cal H}_{\rm BH}^{\rm (distant)}$ in Eq.~(\ref{eq:H-distant}):
\begin{equation}
  \ket{\phi} \in {\cal H}_{\rm BH}^{\rm (distant)}.
\label{eq:BH-before}
\end{equation}
Now, suppose we evolve the state $\ket{\phi}$ back in time to an early 
time $t_0$, perform a boost so that $p$ crosses the horizon at some 
time between $t_0$ and $t$, and then evolve $\ket{\phi}$ forward in time 
to the same $t$.  Under this reference frame change, $\ket{\phi}$ is 
transformed into a state that is not in ${\cal H}_{\rm BH}^{\rm (distant)}$ 
but in ${\cal H}_{\rm BH}^{\rm (falling)}$ in Eq.~(\ref{eq:H-falling}):
\begin{equation}
  \ket{\phi} \quad
  \stackrel{\mbox{\footnotesize boost $B$ at $t_0$}}{\longrightarrow}
  \quad U_B \ket{\phi} \in {\cal H}_{\rm BH}^{\rm (falling)},
\label{eq:BH-after}
\end{equation}
so that the degrees of freedom associated with the horizon, 
${\cal H}_{\rm horizon}$, are mapped into those with the inside 
spacetime, ${\cal H}_{\rm inside}$, {\it which did not exist in the 
${\cal H}_{\cal M}$ containing $\ket{\phi}$ before the transformation}. 
This mapping should be one-to-one if $U_B$ is unitary, which is possible 
because the holographic principle ensures ${\rm dim}\,{\cal H}_{\rm horizon} 
= {\rm dim}\,{\cal H}_{\rm inside}$.  (Here, we have assumed that the 
time $t$ is before $p$ hits the singularity in the boosted frame.) 
Note that both ${\cal H}_{\rm BH}^{\rm (distant)}$ and ${\cal H}_{\rm 
BH}^{\rm (falling)}$ are contained in the full Hilbert space ${\cal H}$ 
as {\it separate} components:
\begin{equation}
  {\cal H} = \cdots \oplus {\cal H}_{\rm BH}^{\rm (distant)} 
    \oplus {\cal H}_{\rm BH}^{\rm (falling)} \oplus \cdots,
\label{eq:H-contain}
\end{equation}
since ${\cal H}$ contains all the possible semi-classical geometries as 
viewed from a local Lorentz frame of $p$.  Thus the transformation at $t$ 
corresponding to the reference frame change $B$ at $t_0$ is represented 
within ${\cal H}$, as it should be.

The statement in Eq.~(\ref{eq:BH-after}) is nothing but black hole 
complementarity.  This, therefore, leads to the following picture. 
{\it Black hole complementarity (or more generally, horizon complementarity) 
arises because changes of the reference frame are represented in 
the Hilbert space of Eq.~(\ref{eq:ST-H}), which contains components 
${\cal H}_{\cal M}$ that are defined only in restricted spacetime 
regions because of the existence of horizons.}  These changes in 
general transform degrees of freedom associated with spacetime 
to those with a horizon, or vice versa---{\it the concept of spacetime 
depends on the reference frame}.

The transformation discussed here can be viewed as an extension of the 
Lorentz/Poincar\'{e} transformation in the quantum gravitational context, 
as the Lorentz transformation is viewed as an extension of the Galilean 
transformation.  For a given $t$, the transformation here is specified 
by $(d-1)(d+2)/2$ parameters:\ $r_i, \eta_i, \theta_{[ij]}$.  In the 
limit $G_N \rightarrow 0$, where relative accelerations between all 
families of geodesics vanish, the transformation is reduced to an 
element of the $(d-1)(d+2)/2$ parameter subset of standard Poincar\'{e} 
transformations consisting of spatial translations, rotations and 
boosts.  Time translation also arises from invariance under a shift 
of the origin of $t$ in the proper time parameterization.  The set 
of these transformations, therefore, is reduced to the standard 
Poincar\'{e} transformations in the limit $G_N \rightarrow 0$.

This is very much analogous to the fact that the standard Lorentz 
transformation is reduced to the Galilean transformation in the limit 
$c \rightarrow \infty$, where $c$ is the speed of light.  In the Galilean 
transformation a change of the reference frame leads only to a constant 
shift of all the velocities, while in the Lorentz transformation it 
also alters temporal and spatial lengths (time dilation and Lorentz 
contraction) and makes the concept of simultaneity relative.  With 
gravity, a change of the reference frame makes even the concept of 
spacetime relative.  The trend is consistent---as we ``turn on'' 
fundamental constants in nature ($c = \infty \rightarrow \mbox{finite}$ 
and $G_N = 0 \rightarrow \mbox{finite}$), physics becomes more and more 
``relative,'' i.e.\ the description of the same physical system from 
different reference frames differ more.  The transformations described 
here (together with time translation) provide the extension of the 
Galilean group with $c$, $G_N$, and $\hbar$ all finite.

\section{Hilbert Space for Quantum Gravity}
\label{sec:Hilbert-QG}

Here we discuss the full Hilbert space for quantum gravity.  We argue 
that it contains an infinite number of ``intrinsically quantum mechanical'' 
states associated with spacetime singularities, which do not admit any 
classical interpretation.  The evolution of the multiverse state is 
unitary in this full Hilbert space, and an arbitrary change of the 
reference frame is represented as a unitary transformation in that 
space.  The ultimate fate of the multiverse state is a superposition 
of supersymmetric Minkowski and singularity states, composing the ``heat 
death of the multiverse.''  This implies that the Hilbert space dimension 
of the quantum universe is infinite, explaining the fact that we observe 
an ordered world obeying consistent laws of physics.

\subsection{Meaning of spacetime singularities}
\label{subsec:singularities}

What happens if a component of Eq.~(\ref{eq:Psi-multiverse}) hits a 
spacetime singularity at some time $t_s$?  We conjecture that it should 
be dropped from physical considerations after time $t_s$.  This is 
motivated by the following independent arguments~\cite{Nomura:2011dt}:
\begin{itemize}
\item
{\it The covariant entropy bound does not count the degrees of freedom 
that have hit a singularity.}  Imagine sending a light sheet inwards from 
a black hole horizon $H$.  The degrees of freedom swiped by the light 
sheet are then bounded by the area of the horizon ${\cal A}_H/4$. 
The entropy bound, however, does not limit the amount of information 
that hits the singularity before being swiped by the light sheet.  The 
simplest interpretation of this fact is that degrees of freedom that 
hit the singularity disappear from the spacetime.
\item
{\it Super-Planckian physics does not have degrees of freedom as 
suggested by field theory.}  Consider a black hole (or de~Sitter) 
horizon.  Because of the blueshift, the local temperature at the 
mathematical horizon formally diverges.  In fact, the physical horizon, 
from which Hawking radiation arises, is a Planckian distance away from 
the mathematical horizon, as suggested by the fact that Bekenstein-Hawking 
(Gibbons-Hawking) entropy is saturated by thermal entropy outside (inside) 
this ``stretched'' horizon.  This suggests that super-Planckian physics 
does not have degrees of freedom as indicated by field theory.  And 
spacetime singularities are precisely the regions where the curvature 
is super-Planckian.
\end{itemize}

What does dropping from consideration mean?  In Ref.~\cite{Nomura:2011dt}, 
it was postulated that a component that hits a singularity at time $t_s$ 
is simply eliminated from the state $\ket{\Psi(t)}$ at that time, $t = t_s$. 
This, however, leads to the following paradoxical situations.  What 
if $\ket{\Psi(t)}$ contains only components that hit singularities 
in the future?  This must violate unitarity, and therefore so does 
any state containing them.  Furthermore, even if we accepted unitarity 
violation, systems having only de~Sitter and anti de~Sitter vacua would 
allow for (a constant fraction of) $\ket{\Psi(t)}$ to stay in a de~Sitter 
phase at $t \rightarrow \infty$, since components tunneled into the anti 
de~Sitter vacuum keep disappearing.  This is extremely counter-intuitive.

We are therefore led to the following picture.%
\footnote{This picture has been developed after discussions with Alan 
 Guth.  I thank him for his insightful suggestions.}
The components that hit spacetime singularities keep existing in 
$\ket{\Psi(t)}$, but they are no longer extracted by projection operators 
${\cal O}_A$ or ${\cal O}_{A \cap B}$ appearing in the probability formula 
because these states cannot be interpreted as those associated with 
(semi-)classical spacetime.  This implies that the full Hilbert space 
for quantum gravity must contain these ``intrinsically quantum mechanical'' 
states, associated with singularities, in addition to those discussed 
in Section~\ref{subsec:Hilbert}:
\begin{equation}
  {\cal H}_{\rm QG} = {\cal H} \oplus {\cal H}_{\rm sing},
\label{eq:QG-H}
\end{equation}
where ${\cal H}$ is the component that allows for spacetime interpretation, 
Eq.~(\ref{eq:ST-H}), while ${\cal H}_{\rm sing}$ is the one that does 
not.  Note that this treatment of ``dropping'' is different from simply 
eliminating the relevant components.  For example, a de~Sitter phase 
can now disappear by tunneling purely into an anti de~Sitter vacuum.

We assume that the evolution of the multiverse state $\ket{\Psi(t)}$ 
is unitary in ${\cal H}_{\rm QG}$, and that any change of the reference 
frame at an arbitrary time $t$ is represented by a unitary transformation 
in this Hilbert space.  According to the current understanding of string 
theory, a full quantum gravitational theory possesses many de~Sitter, 
anti de~Sitter, and Minkowski vacua.  In particular, it possesses exactly 
supersymmetric Minkowski vacua, which are absolutely stable due to the 
positive energy theorem~\cite{Weinberg:1982id}.  This implies that the 
dimension of ${\cal H}$ is infinite, since the Hilbert space dimension 
of (stable) Minkowski space is infinite:
\begin{equation}
  {\rm dim}({\cal H}_{\rm Minkowski}) = \infty
\quad\Longrightarrow\quad
  {\rm dim}({\cal H}) = \infty.
\label{eq:dim-ST-inf}
\end{equation}
What about ${\cal H}_{\rm sing}$?  Consider a set of states that hit 
singularities at some late time.  In the eternally inflating multiverse, 
these states can be mapped into those that evolve into stable 
supersymmetric Minkowski states, by appropriate boost transformations. 
This suggests that
\begin{equation}
  {\rm dim}({\cal H}_{\rm sing}) = \infty,
\label{eq:dim-sing}
\end{equation}
which makes it possible that components that hit singularities do never 
return to states in ${\cal H}$, associated with spacetime.  Namely, stable 
Minkowski and anti de~Sitter vacua can act as ``sinks'' in the landscape, 
despite the fact that the evolution of the multiverse state is unitary 
(at least after the ``birth'').

\subsection{The heat death of the multiverse}
\label{subsec:infinite}

What is the ultimate fate of the multiverse state?  Starting from a generic 
eternally inflating state, the coefficients of any components in unstable 
vacua will eventually decay.  In the string landscape, we expect that all 
the de~Sitter as well as non-supersymmetric Minkowski vacua decay into 
some lower energy vacua, given that there is a huge number of possible 
decay channels.  The multiverse state, therefore, will asymptotically 
become a superposition of supersymmetric Minkowski and singularity states:
\begin{equation}
  \ket{\Psi(t)} \,\,\stackrel{t \rightarrow \infty}{\longrightarrow}\,\, 
    \sum_i a_i(t) \ket{\mbox{supersymmetric Minkowski world $i$}}
  \,+\, \sum_j b_j(t) \ket{\mbox{``singularity world'' $j$}},
\label{eq:asympt}
\end{equation}
where the first sum runs over components with varying matter content, 
spatial dimensions, and the amount of supersymmetries, while the second 
sum contains terms associated with black hole and big crunch singularities. 
In fact, this is simply a consequence of the second law of thermodynamics, 
given that the Hilbert space dimensions of Minkowski and singularity 
worlds are infinite, Eqs.~(\ref{eq:dim-ST-inf},~\ref{eq:dim-sing}). 
In other words, the coarse-grained entropy of the multiverse diverges 
in the asymptotic future:
\begin{equation}
  S_{\rm final} \equiv S_{\rm multiverse}(t \rightarrow \infty) = \infty.
\label{eq:S-final}
\end{equation}

The initial state of the multiverse must be given by some theory external 
to the current framework.  We do not address this issue further here, 
although some possibilities were discussed in Ref.~\cite{Nomura:2011dt}. 
We may naturally speculate that the coarse-grained entropy of the initial 
state is small, e.g.\ $S_{\rm init} \sim O(1)$.  This is, however, not 
required by observation~\cite{Bousso}.

In any event, the fact that $S_{\rm final} = \infty$, i.e.\ the final 
state Hilbert space has an infinite dimensionality, plays a crucial role 
in explaining the basic fact that we perceive an ordered, regular world 
obeying the laws of physics, as discussed in Section~\ref{subsec:inf-needed}. 
Indeed, {\it the ultimate future of the multiverse is the ``heat death,''} 
as represented in Eq.~(\ref{eq:asympt}).

\section{Probabilities in the Quantum Multiverse}
\label{sec:measure}

We now discuss the probability formula in the eternally inflating multiverse. 
This provides a device through which we can relate the ``whole reality'' 
in the multiverse state $\ket{\Psi(t)}$ with our own experience, i.e.\ 
what we---physical objects within $\ket{\Psi(t)}$---may observe.  The 
resulting formula can be applied to answering questions both regarding 
global properties of the universe and outcomes of particular experiments; 
in particular, it reduces to the standard Born rule in a setup corresponding 
to a usual terrestrial experiment.  This, therefore, provides complete 
unification of the two concepts:\ the eternally inflating multiverse 
and many worlds in quantum mechanics~\cite{Nomura:2011dt}.

\subsection{The (extended) Born rule}
\label{subsec:Born}

Having understood how the state $\ket{\Psi(t)}$ is defined, the 
probability formula can be obtained following the earlier discussions 
in Sections~\ref{sec:interpret}~--~\ref{sec:QM-classical}.  An important 
new point here is that the ``time'' $t$ in quantum gravity is simply 
an auxiliary parameter introduced to describe the ``evolution'' of the 
state, exactly like a variable $t$ used in a parametric representation 
of a curve on a plane, $(x(t), y(t))$.  The physical information is 
only in {\it correlations} between events, like correlations between 
$x$ and $y$ in the case of a curve on a plane~\cite{DeWitt:1967yk}. 
Specifically, time evolution of a physical quantity $X$ is nothing more 
than a correlation between $X$ and a quantity that can play the role 
of time, such as the location of the hands of a clock or the average 
temperature of the cosmic microwave background in our universe.

Any physical question can then be phrased as:\ given condition $A$ 
we specify, what is the probability for an event $B$ to occur? 
For example, one can specify a certain ``premeasurement'' situation 
$A_{\rm pre}$ (e.g.\ the configuration of an experimental apparatus 
and the state of an experimenter before measurement) as well as a 
``postmeasurement'' situation $A_{\rm post}$ (e.g.\ those after the 
measurement but without specifying outcome) as $A = \{ A_{\rm pre}, 
A_{\rm post} \}$, and then ask the probability of a particular result 
$B$ (specified, e.g., by a physical configuration of the pointer of 
the apparatus in $A_{\rm post}$) to be obtained.  The information 
about real, physical time is included in conditions $A$ and $B$ through 
specifications of any non-static observables.  The relevant probability 
$P(B|A)$ is then given by
\begin{equation}
  P(B|A) = \frac{\int\!\!\!\int\!dt_1 dt_2 \bra{\Psi(0)} U(0,t_1)\, 
      {\cal O}_{A_{\rm pre}}\, U(t_1,t_2)\, {\cal O}_{A_{\rm post} \cap B}\, 
      U(t_2,t_1)\, {\cal O}_{A_{\rm pre}}\, U(t_1,0) \ket{\Psi(0)}}
    {\int\!\!\!\int\!dt_1 dt_2 \bra{\Psi(0)} U(0,t_1)\, 
      {\cal O}_{A_{\rm pre}}\, U(t_1,t_2)\, {\cal O}_{A_{\rm post}}\, 
      U(t_2,t_1)\, {\cal O}_{A_{\rm pre}}\, U(t_1,0) \ket{\Psi(0)}}.
\label{eq:prob-final}
\end{equation}
Here, $U(t_1,t_2) = e^{-iH(t_1-t_2)}$ is the ``time evolution'' operator 
with $H$ being the Hamiltonian for the entire system, and ${\cal O}_X$ 
is the operator projecting onto states consistent with condition $X$:
\begin{equation}
  {\cal O}_X = \sum_{i \in X} \ket{\alpha_i} \bra{\alpha_i},
\label{eq:O_A-O_B}
\end{equation}
where $\ket{\alpha_i}$ are the classical states discussed in 
Section~\ref{subsec:questions}, and $i \in X$ implies that the sum 
is taken for the configurations that satisfy condition $X$.

Note that since we have already fixed a reference frame, conditions 
$A_{\rm pre}$ and $A_{\rm post}$ in general must involve specifications 
of ranges of location and velocity for physical objects {\it with respect 
to the origin of the coordinates $p$} (in addition to those of physical 
times made through configurations of non-static objects, e.g.\ the hands 
of a clock or the status of an experimenter).  This is important for 
the uniqueness of the framework, eliminating the ambiguity associated 
with how these objects must be counted.  Of course, there is still a 
freedom in specifying in what state the objects must be in $A_{\rm pre}$; 
for example, we could put them at $p$ or some other point at rest, or 
could specify a phase space region in which they must be.  But this 
is the freedom of questions one may ask, and not that of the framework 
itself.  (And the final answer does not depend on the location/velocity 
of reference point $p$, i.e.\ the overall relative location/velocity 
between $p$ and the specified configurations in $A_{\rm pre}$ and 
$A_{\rm post}$, if the multiverse state $\ket{\Psi(t)}$ as a whole 
is invariant under the corresponding reference frame changes.)

Equation~(\ref{eq:prob-final}) is our final formula for the probabilities. 
The integrations over ``time'' $t$ are taken from $t = t_0$, where the 
initial condition for $\ket{\Psi(t)}$ is specified, to $t = \infty$, which 
arise because conditions $A_{\rm pre}$ and $A_{\rm post}$ may be satisfied 
at any values of $t$ (denoted by $t_1$ and $t_2$ in the equation).  This, 
together with appropriate transformations of $H$, ensures that $P(B|A)$ 
is invariant under reparameterization of $t$, as required by general 
covariance.  If conditions $A_{\rm pre}$ and $A_{\rm post}$ are about 
configurations at the same ``time,'' i.e.\ if we ask the question 
like ``given what we know about our past light cone, $A$, what is the 
probability of that {\it same} light cone to have properties $B$ as 
well?,'' then
\begin{equation}
\begin{array}{lcl}
  {\cal O}_{A_{\rm pre}}\, U(t_1,t_2) \,{\cal O}_{A_{\rm post}}\, 
    U(t_2,t_1) \,{\cal O}_{A_{\rm pre}} 
  & \stackrel{t_1=t_2}{\longrightarrow} &
  {\cal O}_{A_{\rm pre}}\, {\cal O}_{A_{\rm post}}\, 
    {\cal O}_{A_{\rm pre}} = {\cal O}_A,
\\
  {\cal O}_{A_{\rm pre}}\, U(t_1,t_2) \,{\cal O}_{A_{\rm post} \cap B}\, 
    U(t_2,t_1) \,{\cal O}_{A_{\rm pre}} 
  & \stackrel{t_1=t_2}{\longrightarrow} &
  {\cal O}_{A_{\rm pre}}\, {\cal O}_{A_{\rm post} \cap B}\, 
    {\cal O}_{A_{\rm pre}} = {\cal O}_{A \cap B},
\end{array}
\label{eq:t_1=t_2}
\end{equation}
so that Eq.~(\ref{eq:prob-final}) is reduced to the formula of 
Eq.~(\ref{eq:probability-AB}).  (One can see this explicitly by inserting 
$\delta(t_1-t_2)$ both in the numerator and denominator and using 
$\ket{\Psi(t)} = U(t,0) \ket{\Psi(0)}$.)%
\footnote{The formula in Eq.~(\ref{eq:prob-final}) treats the state 
 between $t_1$ and $t_2$ as that evolved from ${\cal O}_{A_{\rm pre}}\! 
 \ket{\Psi(t_1)}$, not $\ket{\Psi(t_1)}$.  This is justified under the 
 current setup, i.e.\ a generic initial state $\ket{\Psi(t_0)}$ evolving 
 as in Eq.~(\ref{eq:asympt}), but not in the case where the possibility 
 of recoherence cannot be ignored even for macroscopic objects.  In such 
 a case, the formula in Eq.~(\ref{eq:probability-AB}) must be used, which 
 applies in any quantum mechanical system (and allows for answering any 
 physical questions, although it requires some elaboration to apply to 
 questions like the ones asked in Eq.~(\ref{eq:prob-final}).)}

In many practical situations, we do not know the exact (multiverse) 
state, so that the system is described by a density matrix:\ $\rho(t) 
\equiv \sum_i p_i \ket{\Psi_i(t)} \bra{\Psi_i(t)}$ with $\sum_i p_i = 1$. 
In this case, Eq.~(\ref{eq:prob-final}) is straightforwardly extended to
\begin{equation}
  P(B|A) = \frac{\int\!\!\!\int\!dt_1 dt_2\, {\rm Tr}\left[ \rho(0)\, 
    U(0,t_1)\, {\cal O}_{A_{\rm pre}}\, U(t_1,t_2)\, 
    {\cal O}_{A_{\rm post} \cap B}\, U(t_2,t_1)\, {\cal O}_{A_{\rm pre}}\, 
    U(t_1,0) \right]}
  {\int\!\!\!\int\!dt_1 dt_2\, {\rm Tr}\left[ \rho(0)\, 
    U(0,t_1)\, {\cal O}_{A_{\rm pre}}\, U(t_1,t_2)\, 
    {\cal O}_{A_{\rm post}}\, U(t_2,t_1)\, {\cal O}_{A_{\rm pre}}\, 
    U(t_1,0) \right]}.
\label{eq:prob-final-rho}
\end{equation}
This formula is also applicable for a reduced density matrix, obtained 
by tracing out some of the degrees of freedom, with the understanding 
that ${\cal O}_X$ act on degrees of freedom that are not traced 
out.  An important example includes the bulk density matrix, which 
is obtained by tracing out horizon (and possibly, singularity) degrees 
of freedom~\cite{Nomura:2011dt}
\begin{equation}
  \rho_{\rm bulk}(t) = {\rm Tr}_{\rm horizon}\, \rho(t).
\label{eq:rho-red_bulk}
\end{equation}
Since the evolution of $\rho_{\rm bulk}(t)$ can be determined by 
semi-classical calculations, this allows us to make predictions/postdictions 
in the multiverse without knowing full quantum gravity (at the cost of 
unitarity in processes involving horizons/singularities).

\subsection{Unification of the eternally inflating multiverse and many 
 worlds in quantum mechanics}

The probability formula of Eq.~(\ref{eq:prob-final}), or 
Eq.~(\ref{eq:prob-final-rho}), can be applied to any physical 
questions, from the smallest possible (Planck length) to the largest 
possible (multiverse) scales.

Under the usual situation of a terrestrial experiment, the formula is 
reduced to the standard Born rule.  This can be seen by isolating the 
degrees of freedom relevant to the experiment (which may involve the 
apparatus and/or experimenter, in addition to the system to be measured). 
Suppose $O_{A_{\rm pre}}$ acts on these degrees of freedom and selects 
a particular premeasurement situation $A_{\rm pre}$, which is realized 
in components of $\ket{\Psi(t)}$ at multiple values of $t = \hat{t}_i$ 
($i=1,2,\cdots$):
\begin{equation}
  O_{A_{\rm pre}} U(t_1,0) \ket{\Psi(0)} 
  = \sum_i c_i \ket{\phi(\hat{t}_i)} \otimes 
      \bigl| \hat{\Psi}_i(\hat{t}_i) \bigr>\, \delta(t_1-\hat{t}_i),
\label{eq:QM-initial}
\end{equation}
where $\ket{\phi(\hat{t}_i)}$ represents the degrees of freedom relevant 
to the experiment in a configuration selected by $A_{\rm pre}$, and 
$\bigl| \hat{\Psi}_i(\hat{t}_i) \bigr>$ the other degrees of freedom 
in the relevant component of $\ket{\Psi(t)}$ at $t = \hat{t}_i$.  In 
the limit that $A_{\rm pre}$ selects the initial experimental setup 
infinitely accurately, which we are considering here for simplicity, 
the initial state for the experiment is
\begin{equation}
  \ket{\phi(\hat{t}_1)} = \ket{\phi(\hat{t}_2)} = \cdots 
  \equiv \ket{\phi(t_{\rm before})}.
\label{eq:exp-initial}
\end{equation}
Then, assuming that $O_{A_{\rm post}}$, which selects a particular 
postmeasurement situation, acts on the same degrees of freedom as 
$O_{A_{\rm pre}}$, and that the effect of the other degrees of freedom 
on the experimental system is negligible, we can write as
\begin{equation}
  O_{A_{\rm post}} \hat{U}(t_2,t_{\rm before}) \ket{\phi(t_{\rm before})} 
  = c \ket{\phi(t_{\rm after})} \delta(t_2 - t_{\rm after}) + \cdots,
\label{eq:exp-after}
\end{equation}
where $\hat{U}(t_1,t_2)$ is a factor in $U(t_1,t_2)$ acting only on the 
experimental degrees of freedom, and $t_{\rm after}$ is the smallest 
value of  $t_2$ consistent with the projection $O_{A_{\rm post}}$, which 
completely dominates the right-hand side.  (In the usual setup of a 
terrestrial experiment, the coefficients of the terms in $\cdots$ are 
exponentially smaller than $c$.)  Rewriting ${\cal O}_{A_{\rm post}}$ 
and ${\cal O}_{A_{\rm post} \cap B}$ as ${\cal O}_{\rm obs}$ and 
${\cal O}_{{\rm obs} \cap \alpha}$, respectively, Eq.~(\ref{eq:prob-final}) 
then gives the probability of obtaining outcome $\alpha$
\begin{equation}
  P(\alpha|{\rm obs}) = \frac{\bra{\phi(t_{\rm after})} 
    {\cal O}_{{\rm obs} \cap \alpha} \ket{\phi(t_{\rm after})}}
  {\bra{\phi(t_{\rm after})} 
    {\cal O}_{\rm obs} \left| \phi(t_{\rm after}) \right>}.
\label{eq:prob-Born}
\end{equation}
(All the other factors cancel between the numerator and the denominator.) 
This is nothing but the usual Born rule.

At scales of our everyday life, the formulae of 
Eqs.~(\ref{eq:prob-final},~\ref{eq:prob-final-rho}) should 
(approximately) reproduce Newtonian dynamics.  To see how the standard 
picture of time evolution arises, consider following the motion of a 
baseball (with position ${\bf x}$ and velocity ${\bf v}$) as a function 
of the hands of a clock $\tau$, in the absence of any force acting on it. 
In this case, we should take
\begin{eqnarray}
  {\cal O}_{A_{\rm pre}} &=& \ket{{\bf x},{\bf v}}\bra{{\bf x},{\bf v}} 
    \otimes \ket{\tau}\bra{\tau},
\label{eq:O_Apre-ball}\\
  {\cal O}_{A_{\rm post}} &=& {\bf 1} \otimes \ket{\tau+d\tau}\bra{\tau+d\tau},
\label{eq:O_Apost-ball}\\
  {\cal O}_{A_{\rm post} \cap B} &=& 
    \ket{{\bf x'},{\bf v'}}\bra{{\bf x'},{\bf v'}} 
    \otimes \ket{\tau+d\tau}\bra{\tau+d\tau},
\label{eq:O_B-ball}
\end{eqnarray}
(with some widths around these values), and the probability $P(B|A)$ 
will then be peaked at
\begin{eqnarray}
  {\bf x'} &=& {\bf x} + {\bf v}\,d\tau,
\label{eq:Newton-1}\\
  {\bf v'} &=& {\bf v},
\label{eq:Newton-2}
\end{eqnarray}
as indicated by the Newtonian mechanics.  Note that $\tau$ here represents 
the physical location of the hands of the clock, and {\it not} the ``time'' 
variable $t$, which is integrated out to obtain the probability.

The formulae of Eqs.~(\ref{eq:prob-final},~\ref{eq:prob-final-rho}) 
can also be used to answer questions regarding global properties of 
our universe.  To predict/postdict physical parameters $x$, for example, 
we need to choose $A$ to select the situation of making a measurement 
of $x$.  We can then use various different values (ranges) of $x$ for 
$B$, to obtain the probability distribution $P(x)$.  Despite the fact 
that the $t$ integrals run to $\infty$, the resulting $P(B|A)$ is 
well-defined, since $\ket{\Psi(t)}$ is continually ``diluted'' into 
supersymmetric Minkowski and singularity states~\cite{Nomura:2011dt}. 
The procedure to make predictions/postdictions in this way was discussed 
in Ref.~\cite{Larsen:2011mi}, where the probability distribution of 
the vacuum energy, $x = \rho_\Lambda$, was computed and shown to 
agree with observation at an order of magnitude level.

It is striking that the simple, basic formalism developed here applies 
to physics at all scales.  In particular, the single probability 
formula Eq.~(\ref{eq:prob-final}) (or Eq.~(\ref{eq:prob-final-rho})) can 
be used to answer any physical questions, given a state $\ket{\Psi(t)}$ 
(or $\rho(t)$).  This, therefore, provides complete unification of 
the eternally inflating multiverse and many worlds in quantum mechanics. 
These two are really the same thing---they simply refer to the same 
phenomenon occurring at (vastly) different scales.

\section{Summary}
\label{sec:summary}

An essential feature of quantum mechanics is that information is fragile---as 
is well known, a generic measurement destroys the state of the system 
after the measurement is performed.  Indeed, quantum information cannot 
be faithfully duplicated:\ the exact identification of a single state is 
not possible without having any prior knowledge of the state.  Moreover, 
quantum information transmitted through physical processes will in 
general become non-local, encoded in the entanglement structure of 
a resulting quantum state.

Despite this intrinsically non-local nature of quantum states, however, 
the dynamics {\it is} local.  Specifically, the time evolution operator 
takes a special form such that the concept of locality can be defined 
in spacetime.  This feature allows for a limited set of information 
(among the full quantum information) to be copiously duplicated, i.e.\ 
``amplified,'' and it is (only) this information that we can meaningfully 
store, compare, and handle.  Because of the structure of the evolution 
operator, the relevant information is associated with well-defined 
classical configurations in phase space, at least for macroscopic systems.

In a system with gravity, the whole picture is more subtle, since if we 
choose wrong quantization hypersurfaces, then spacetime locality is not 
manifest {\it even at distances much larger than the quantum gravity scale}. 
We argued, however, that spacetime locality can be preserved if we define 
quantum states in restricted spacetime regions:\ in and on (stretched) 
apparent horizons {\it as viewed from a local Lorentz frame of a fixed 
spatial point $p$}.  This can be viewed as a ``unitary gauge for quantum 
gravity,'' on which our intuition should be based.  By appropriately 
limiting the dimensions of the Hilbert subspaces corresponding to a fixed 
semi-classical geometry, all the redundancies/overcountings associated 
with a general relativistic, global spacetime description of nature are 
fixed/eliminated.  These include general covariance, global overcounting 
related to complementarity, and local overcounting implied by the 
holographic principle.

The need for fixing a reference frame in describing the gravitational 
system quantum mechanically was emphasized.  We identified the 
transformation associated with changes of the reference frame, which 
is specified by $d(d+1)/2$ parameters in $d$ spacetime dimensions. 
This transformation is the origin of horizon complementarity as well 
as the ``observer dependence'' of horizons {\it and spacetime}. 
The transformation is reduced to the standard Poincar\'{e} transformation 
in the limit $G_N \rightarrow 0$, so it can naturally be regarded as an 
extension of the Poincar\'{e} transformation in the quantum gravitational 
context.  This is much like that the standard Lorentz transformation 
is regarded as an extension of the Galilean transformation, where the 
former is reduced to the latter in the limit $c \rightarrow \infty$.

It is remarkable that the simple framework described in this paper 
is applicable to physics at all scales, from the smallest (Planck length) 
to the largest (multiverse).  Indeed, it is quite striking that quantum 
mechanics does not need any modification to be applied to phenomena at 
such vastly different scales.  In the 20th century, we have witnessed 
the tremendous success of quantum mechanics, following its birth at the 
beginning.  In the early 21st century, quantum mechanics seems still 
giving us an opportunity to explore deep facts about nature, such as 
spacetime and gravity.  Does quantum mechanics break down at some point? 
We don't know.  But perhaps, the beginning of the multiverse might 
provide one.

\section*{Acknowledgments}

I am grateful for useful discussions with Alan Guth, Grant Larsen, and 
I-Sheng Yang.  This work was supported in part by the Director, Office 
of Science, Office of High Energy and Nuclear Physics, of the US Department 
of Energy under Contract DE-AC02-05CH11231, and in part by the National 
Science Foundation under grant PHY-0855653.

\appendix

\section{Quantum Measurement:\ Dissipation of Coherence into 
 (Infinitely) Large Hilbert Space}
\label{app:dynamics}

The picture of quantum measurement presented in this paper involves 
several stages of ``dissipation of coherence'' into larger Hilbert 
spaces:\ from microscopic to macroscopic, eventually to a supersymmetric 
Minkowski or singularity world.  At each stage, a superposition of the 
smaller system in some basis is translated into a superposition of the 
larger system.  In the early stages of this process, i.e.\ when the 
systems involved are small, the selected bases depend on the details 
of the setup, as the standard analysis of decoherence shows.  However, 
at the later stages, where the relevant systems become larger, the 
appropriate bases quickly approach the classical state basis, as 
discussed in Section~\ref{sec:QM-classical}.  In those stages, information 
associated with the selected basis of the original system is tremendously 
amplified {\it in each term}, producing an effective classical world. 
These classical worlds, represented by various terms in the full 
multiverse state, then evolve independently and eventually become 
different supersymmetric Minkowski (or singularity) worlds.  Since the 
Hilbert space dimension of Minkowski states is infinite, the resulting 
worlds do not recohere.

The picture described above provides an ultimate answer to the question of 
basis selection in quantum measurement.  Here we focus only on one classic 
example:\ why the apparatus $\apparatusalpha$ measures the spin in the 
$z$ direction and not in the $x$ direction.  The question is typically 
phrased in the following form.  Suppose the apparatus interacted with 
the spin $(\ket{\uparrow} + \ket{\downarrow})/\sqrt{2}$.  Then the 
combined apparatus-spin system after the measurement can be written 
either as
\begin{equation}
  \ket{\Psi} = 
  \frac{1}{\sqrt{2}} \ket{\uparrow} \bigl|\apparatusup\bigr> 
  + \frac{1}{\sqrt{2}} \ket{\downarrow} \bigl|\apparatusdown\bigr> 
\label{eq:after-1}
\end{equation}
or
\begin{equation}
  \ket{\Psi} = 
  \frac{1}{2\sqrt{2}} (\ket{\uparrow} + \ket{\downarrow}) 
    \bigl( \bigl|\apparatusup\bigr> + \bigl|\apparatusdown\bigr> \bigr) 
  + \frac{1}{\sqrt{2}} (\ket{\uparrow} - \ket{\downarrow}) 
    \bigl( \bigl|\apparatusup\bigr> - \bigl|\apparatusdown\bigr> \bigr),
\label{eq:after-2}
\end{equation}
where we have omitted the symbol $\otimes$ for notational simplicity. 
This seems to indicate that there is an ambiguity associated with the 
basis in which the measurement has been performed:\ along the $z$ axis 
(Eq.~(\ref{eq:after-1})) or the $x$ axis (Eq.~(\ref{eq:after-2})).

Since the apparatus is macroscopic, however, further dissipation/amplification 
of coherence occurs in the classical state basis, which is more or less 
a ``smeared locality basis.''  For example, in a complete 
treatment of the measurement including the observer and the desk, 
the state after the measurement is given by Eq.~(\ref{eq:Psi-3}) with 
$c_\uparrow = c_\downarrow = 1/\sqrt{2}$:
\begin{equation}
  \ket{\Psi(t \gg t_{\rm obs})} 
  = \frac{1}{\sqrt{2}} \ket{\uparrow} \bigl|\apparatusup\bigr> 
    \bigl|\desk\bigr> \bigl|\obsafter\brainup\bigr> 
    + \frac{1}{\sqrt{2}} \ket{\downarrow} \bigl|\apparatusdown\bigr> 
    \bigl|\desk\bigr> \bigl|\obsafter\braindown\bigr>.
\label{eq:after-3}
\end{equation}
This is enough to conclude that the experiment measures the spin in the 
$z$ direction, i.e.\ the direction perpendicular to the surface of the 
desk, because the two terms will evolve (eventually) into two different 
supersymmetric Minkowski (or singularity) states, which will never 
recohere.  It is clear that writing the state as in Eq.~(\ref{eq:after-2}),
\begin{eqnarray}
  && \ket{\Psi(t \gg t_{\rm obs})} 
  = \frac{1}{2\sqrt{2}} (\ket{\uparrow} + \ket{\downarrow}) 
    \Bigl( \bigl|\apparatusup\bigr> \bigl|\desk\bigr> 
    \bigl|\obsafter\brainup\bigr> + \bigl|\apparatusdown\bigr> 
    \bigl|\desk\bigr> \bigl|\obsafter\braindown\bigr> \Bigr) 
\nonumber\\
  && \qquad\qquad\qquad\,
    + \frac{1}{2\sqrt{2}} (\ket{\uparrow} - \ket{\downarrow}) 
    \Bigl( \bigl|\apparatusup\bigr> \bigl|\desk\bigr> 
    \bigl|\obsafter\brainup\bigr> - \bigl|\apparatusdown\bigr> 
    \bigl|\desk\bigr> \bigl|\obsafter\braindown\bigr> \Bigr),
\label{eq:after-4}
\end{eqnarray}
does not affect physical predictions.

Of course, we could alternatively measure the spin in the $x$ direction, 
instead of the $z$ direction.  To do so, however, we need to prepare 
a different apparatus (or arrange a different configuration of the detector). 
Representing such an apparatus by a square, the dynamical evolution is now
\begin{equation}
  \frac{1}{\sqrt{2}} (\ket{\uparrow} + \ket{\downarrow}) 
    \bigl|\apparatusxneut\bigr>
\rightarrow
  \frac{1}{\sqrt{2}} (\ket{\uparrow} + \ket{\downarrow}) 
    \bigl|\apparatusxright\bigr>,
\label{eq:meas-x_1}
\end{equation}
or in a more complete description
\begin{eqnarray}
  && \ket{\Psi(t \ll t_{\rm m})} 
  = \frac{1}{\sqrt{2}} (\ket{\uparrow} + \ket{\downarrow}) 
    \bigl|\apparatusxneut\bigr> \bigl|\desk\bigr> \bigl|\obsbefore\bigr>
\label{eq:meas-x_before}\\
  && \!\!\!\!\!\! \longrightarrow
\nonumber\\
  && \ket{\Psi(t \gg t_{\rm obs})} 
  = \frac{1}{\sqrt{2}} (\ket{\uparrow} + \ket{\downarrow}) 
    \bigl|\apparatusxright\bigr> \bigl|\desk\bigr> 
    \bigl|\obsafter\brainright\bigr>.
\label{eq:meas-x_after}
\end{eqnarray}
In this case, Eq.~(\ref{eq:meas-x_after}) itself is the selected, classical 
state basis, corresponding to the outcome of the experiment.

In practice, once coherence is promoted to a sufficiently macroscopic 
level, we can regard components of a state that have different phase 
space configurations as {\it different worlds}, since the probability 
for them to recohere is infinitesimally small.  As we have seen, this is 
a consequence of spacetime locality---the fact that the time evolution 
operator takes a special form of Eq.~(\ref{eq:QFT-local})---as well as 
the infinite dimensionality of the Hilbert space of the quantum universe. 
Ultimately, it is these features that are responsible for the appearance 
of (apparently classical, and ordered) many worlds in the quantum 
mechanical universe.

\end{document}